\PassOptionsToPackage{unicode}{hyperref}
\PassOptionsToPackage{hyphens}{url}
\documentclass[14pt]{article}
\usepackage{amsmath,amssymb}
\usepackage{lmodern}
\usepackage{makecell}  
\usepackage{iftex}
\ifPDFTeX
  \usepackage[T1]{fontenc}
  \usepackage[utf8]{inputenc}
  \usepackage{textcomp} 
\else 
  \usepackage{unicode-math}
  \defaultfontfeatures{Scale=MatchLowercase}
  \defaultfontfeatures[\rmfamily]{Ligatures=TeX,Scale=1}
\fi
\IfFileExists{upquote.sty}{\usepackage{upquote}}{}
\IfFileExists{microtype.sty}{
  \usepackage[]{microtype}
  \UseMicrotypeSet[protrusion]{basicmath} 
}{}
\makeatletter
\@ifundefined{KOMAClassName}{
  \IfFileExists{parskip.sty}{%
    \usepackage{parskip}
  }{
    \setlength{\parindent}{0pt}
    \setlength{\parskip}{6pt plus 2pt minus 1pt}}
}{
  \KOMAoptions{parskip=half}}
\makeatother
\usepackage{xcolor}
\usepackage{algorithm}
\usepackage{algpseudocode}
\usepackage{longtable,booktabs,array}
\usepackage{calc} 
\usepackage{etoolbox}
\makeatletter
\patchcmd\longtable{\par}{\if@noskipsec\mbox{}\fi\par}{}{}
\makeatother
\IfFileExists{footnotehyper.sty}{\usepackage{footnotehyper}}{\usepackage{footnote}}
\makesavenoteenv{longtable}
\usepackage{graphicx}
\makeatletter
\def\maxwidth{\ifdim\Gin@nat@width>\linewidth\linewidth\else\Gin@nat@width\fi}
\def\maxheight{\ifdim\Gin@nat@height>\textheight\textheight\else\Gin@nat@height\fi}
\makeatother
\setkeys{Gin}{width=\maxwidth,height=\maxheight,keepaspectratio}
\makeatletter
\def\fps@figure{htbp}
\makeatother
\ifLuaTeX
  \usepackage{selnolig}  
\fi
\ifPDFTeX
  \newcommand{\RL}[1]{\beginR #1\endR}

\fi
\IfFileExists{bookmark.sty}{\usepackage{bookmark}}{\usepackage{hyperref}}
\IfFileExists{xurl.sty}{\usepackage{xurl}}{} 
\hypersetup{
  hidelinks,
  pdfcreator={LaTeX via pandoc}}

\author{}
\date{}
\usepackage{geometry}
\usepackage{setspace}
\usepackage{titlesec}
\usepackage{lmodern}
\usepackage{graphicx}
\usepackage{hyperref}

\titleformat{\section}{\large\bfseries}{\thesection}{0.5 em}{}
\titleformat{\subsection}{\normalsize\bfseries}{\thesubsection}{0.5em}{}

\begin{document}

\begin{center}
    \Large \textbf{The Impact of Artificial Intelligence on the Evolution of Digital Education: A Comparative Study of OpenAI Text Generation Tools including ChatGPT, Bing Chat, Bard, and Ernie} \\[1.1 em]
    
\normalsize

\textbf{Negin Yazdani Motlagh}\textsuperscript{1}, 
\textbf{Matin Khajavi}\textsuperscript{2}, 
\textbf{Abbas Sharifi}\textsuperscript{3}, 
\textbf{Mohsen Ahmadi}\textsuperscript{4}

\footnotesize
    \textsuperscript{1}Department of English literature and language St.Mary’s University of San Antonio, Texas, USA\\
    \textsuperscript{2} Foster School of Businesses, University of Washington, Washington, USA\\
    \textsuperscript{3}Department of Civil and Environmental Engineering, Florida International University, Miami, FL, USA\\
    \textsuperscript{4}Department of Electrical Engineering and Computer Science, Florida Atlantic University, FL, USA\\
    \emph{*Corresponding author: mahmadi2021@fau.edu}

\end{center}

\vspace{0.5em}

\textbf{Abstract}

In the digital era, the integration of artificial intelligence (AI) into
education has revolutionized teaching methodologies, curriculum design,
and student engagement, heralding transformative changes. This review
paper explores the rapidly evolving landscape of digital education by
comparing the capabilities and impacts of OpenAI\textquotesingle s text
generation tools, including Bing Chat, Bard, Ernie, and ChatGPT, with a
particular focus on the latter. Framed within a typology that examines
education through the dimensions of system, process, and result, the
paper delves into the diverse applications of AI in education.
Decentralizing global education exemplified by enabling personalized
learning across geographical and economic boundaries has been realized
through AI\textquotesingle s potential to customize curriculums and
document competence-based outcomes digitally. Highlighting
ChatGPT\textquotesingle s unprecedented rise to one million users in
just five days, this study underscores its pivotal role in democratizing
education, fostering self-directed learning, and enhancing student
engagement. However, with this transformative potential comes the risk
of misuse, particularly in the realm of academic integrity, as AI tools
may inadvertently facilitate unethical practices. This paper contrasts
the promises and pitfalls of AI in education, advocating for a balanced
integration of AI tools and traditional pedagogical approaches.
Recommendations include the establishment of ethical guidelines,
curriculum adaptations, and collaborative strategies among stakeholders.
Concluding, this paper provides insights into the implications of AI in
education while emphasizing its critical contributions: an in-depth
comparative analysis of AI tools, a comprehensive exploration of their
ethical implications, and actionable strategies for effectively
integrating AI into digital education.


\section{1.Introduction}

The integration of artificial intelligence (AI) into education has
brought transformative changes, redefining how teaching methodologies,
curriculum planning, and student engagement are approached in the
digital era. AI specifically influences digital education by enabling
personalized learning, automating administrative tasks, and providing
real-time feedback to enhance educational outcomes. This review paper
delves into the rapidly evolving landscape of digital education by
examining the applications, potential, and challenges of AI tools like
ChatGPT in reshaping traditional educational paradigms. Artificial
intelligence can automate various repetitive and tedious tasks in the
workplace, thereby increasing productivity and efficiency. In the realm
of education, students benefit from individualized learning
opportunities, and teachers can employ cutting-edge instructional
methods. Artificial intelligence techniques, such as machine learning
(ML) and deep learning (DL), are considered revolutionary in various
fields, including but not limited to telecommunications, construction,
transportation, healthcare, manufacturing, marketing, and education
{[}1-3{]}. In higher education, artificial intelligence plays a crucial
role by offering customized learning solutions tailored to each
student\textquotesingle s needs, comprehension, learning pace, and
academic goals {[}4{]}. To enhance students\textquotesingle{} learning
experiences, AI-enhanced educational tools may assess their prior
learning habits to identify areas that need improvement and provide them
with the best, most personalized instruction {[}4,5{]}. Additionally, AI
automates routine administrative duties in universities and colleges,
freeing up more time for teaching and research. Due to the COVID-19
epidemic, all academic institutions have adopted digital teaching
strategies {[}6,7{]}, leading educators and students to actively discuss
this transformation and its implications. Artificial intelligence (AI)
has the potential to transform digital learning, improve teaching
methods, and influence the course of digital education {[}8{]}. Digital
education, as the name implies, refers to the seamless integration of
digital technology into the teaching and learning process, utilizing
digital tools in various educational contexts, from traditional
classroom settings to fully online platforms. To address real-world
problems, AI uses intelligent applications within these constraints,
with machine learning (ML), a subset of AI, learning and adjusting from
information and experiences, and deep learning (DL), another subset of
ML, exploring complex patterns similar to human cognitive processes
{[}9-12{]} (see Figure 1). Thus, it is imperative that these components
are thoroughly examined from an educational perspective. While the study
is focused on higher education, it is important to note that AI-powered
systems may present unforeseen challenges that could also be viewed as
opportunities for improvement. In educational settings, generative AI
could be used by students to produce material that appears unique for
their projects, a practice resembling contract cheating. Although the
text appears to be written by the student, it is not. Unlike traditional
contract cheating, generative AI can create this material almost
instantly and often without incurring any costs. Given the correct
prompt, a student may even provide distinct responses to each classmate
with just one click.

\begin{figure}[htbp]
    \centering
    \includegraphics[width=0.9\textwidth]{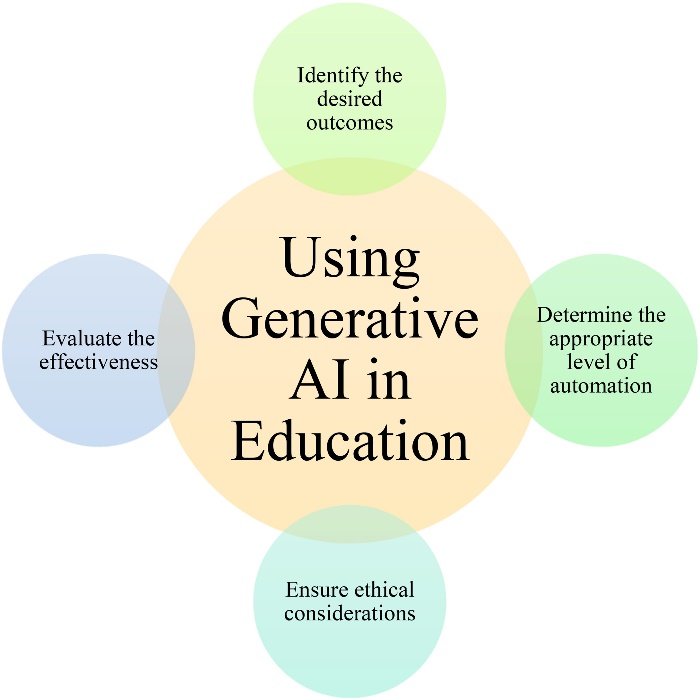}
    \caption{\textbf{}Artificial Intelligence (AI) is known for its wide range of applications, including its growing impact in the field of education.}
    \label{fig:ai_education}
\end{figure}

Automated academic text production is comparable to contract cheating, a
phrase first used by Clarke and Lancaster (2006) {[}10{]} to describe
the practice of students using third parties to perform assignments
dishonestly. Researchers have examined various topics related to
contract cheating, including the reasons students outsource their work
{[}11{]}, detection methods, dangers of exposure to third parties
{[}12{]}, the size of the industry {[}13{]}, and the authors involved
{[}14{]}. Unlike contract cheating, text creation tools do not require
the hiring of a person, allowing students to work independently. The
continual improvement of such tools has been the subject of numerous
technical publications, but they are not included in this study since
they are outside the scope of academic integrity discussion. There was
little research conducted on academic integrity or education prior to
ChatGPT. The study discusses the applications of AI in education. For
example, AI tools like ChatGPT have been used to personalize language
education by providing real-time feedback on grammar and pronunciation,
enabling learners to improve communication skills effectively. This
demonstrates how AI fosters individualized learning experiences that
adapt to students\textquotesingle{} needs and progress. Chrisinger
(2023) {[}15{]} brought to light that existing plagiarism detection
techniques are ineffective when dealing with created material, which is
also a problem in the area of contract fraud. According to Roe and
Perkins (2022), {[}16{]} paraphrasing tools do pose a potential threat
to academic integrity, albeit to a lesser extent. Paraphrasing tools
have been found to confuse originality verification software in previous
studies such as {[}17{]}, and similar findings have also been found in
the related field of essay spinning {[}18{]}. The purpose of this paper is to introduce the educational community to
modern AI text generation tools, such as ChatGPT and other OpenAI tools,
while addressing their potential misuse and impact on academic
integrity. Unauthorized use of these tools may allow students to gain
credit without genuine learning, raising ethical concerns. This paper
explores examples of generated solutions to assignments and explains
techniques, such as systematic repetition of textual output, to mitigate
potential abuse. Furthermore, it provides actionable recommendations for
educators to adapt to the rise of text creation technologies. Employing
case studies and analogies for accessibility, rather than formal
mathematical language, the paper aims to thoroughly examine the
utilization of AI, including machine learning (ML) and deep learning
(DL), in digital education. The study makes two primary contributions:
first, it follows a repeatable and unbiased process to review relevant
literature; and second, it identifies and clarifies key themes related
to the use of AI-driven algorithms in digital education. The research
highlights three significant gaps in the current literature:

\begin{itemize}
\item
  A lack of comparative analysis among major AI text generation tools,
  particularly in their application to digital education.
\item
  Insufficient exploration of ethical implications and academic
  integrity challenges posed by AI tools like ChatGPT.
\item
  Limited evidence-based recommendations for integrating AI tools into
  pedagogical practices.
\end{itemize}

these contributions and gaps, this paper aims to provide educators with
practical insights and strategies for navigating the evolving landscape
of AI in digital education. This study is organized as follows: Section
2 reviews related work, discussing prior studies on the integration of
artificial intelligence (AI) in digital education and its implications.
The methodology used in this study is described in Section 3, which
outlines how AI tools, such as ChatGPT and other text generation
technologies, were selected and evaluated. A comparative analysis of the
tools and their applications in education is presented in Section 4. The
fifth section of the report discusses the findings of the study,
including ethical considerations, academic integrity challenges, and
strategies for integrating artificial intelligence into the educational
system effectively. The final section of the paper summarizes the key
insights, highlights contributions, and proposes future directions for
research in AI-enhanced digital education.

\section{2.Related Work}

Scientists argue that artificial intelligence (AI) is a rapidly
developing technological field with a variety of applications {[}19{]}.
It encompasses many areas of computer science that mimic human cognitive
processes, such as learning, problem-solving, and pattern recognition
{[}20{]}. Due to its adaptability, AI can be used for both serious and
lighthearted activities, impacting daily life and job automation. Some
view AI as a technological advancement that facilitates human activity
and productivity {[}20{]}, while others note that human work may be
negatively affected by AI, despite its advantages. Fitria {[}21{]}
describes artificial intelligence (AI) as a computer-based simulation of
human intelligence that aims to replicate human decision-making,
reasoning, and behavior. This concept is pivotal in advancing
technology, particularly as AI gains popularity across various
industries, with chatbots being a prime example of its practical
application {[}22--24{]}. Chatbots are software programs that utilize
natural language processing (NLP) and deep learning to conduct
text-based discussions {[}24{]}. The NLP is a field of AI focused on
enabling computers to understand, interpret, and respond to human
language in a meaningful way which the Scholars have long been
interested in this topic {[}25{]}, and as computer-human communication
becomes more similar to human interaction, the significance of chatbots
capable of human-like interaction grows. A growing trend in technology
is the use of chatbots to communicate with computers. There has been
increasing interest in using intelligent systems as conversational
agents to facilitate and enhance conversations {[}25-27{]}. These
systems were developed using AI and natural language processing
techniques, and they provide valuable information related to education.
The rapid development of AI has contributed to the growing capabilities
of text generation technologies like GPT-3 and ChatGPT {[}28{]}. ChatGPT
was developed by the San Francisco, California-based company OpenAI. The
release of GPT-3 in 2020 has sparked discussions regarding its potential
and limitations, with applications ranging from summarizing legal
documents to assisting programmers. OpenAI, a non-profit organization
founded by Elon Musk and Sam Altman in 2015 with a \$1 billion
investment, released an enhanced version of GPT-3 called ChatGPT
{[}29{]}. \RL{}Therefore, AI technology, as demonstrated by
OpenAI\textquotesingle s products like DALL-E and GPT, offers a wealth
of benefits to mankind, illuminating the field\textquotesingle s
potential and continued development. In order to facilitate public
collaboration between multiple research groups and individuals, OpenAI
was created. According to the company\textquotesingle s mission
statement, collaboration and openness are key components of the
company\textquotesingle s approach to its patent portfolio and research
discoveries. This strategy prevents behaviors that could jeopardize
safety. One of OpenAI\textquotesingle s products, ChatGPT, predicts the
words and phrases that might be used in a text or discussion.
Essentially, ChatGPT functions as a chatbot, a virtual robot that mimics
human communication. Although still a prototype, this dialog-based AI
chatbot has demonstrated the ability to automatically respond to natural
human language {[}30,31{]}.

Some have speculated that the platform might
replace Google due to its ability to handle a wide range of issues.
Compared to other chatbots, ChatGPT stands out for its ability to
respond to challenging inquiries comprehensively. As a version of
InstructGPT, ChatGPT OpenAI follows instructions and provides detailed
feedback. It can even recall prior inquiries and modify its responses in
response to feedback {[}32{]}. According to Rahman et al. (2017)
{[}33{]}, several cloud-based chatbot services are accessible. Various
studies have shown that chatbots can promote health programs, facilitate
customer service, and serve as interactive information tools. Evidence
supports that chatbots enhance customer experience and can be used by
companies and organizations without specific staff to manage
question-and-answer services. Chatbots\textquotesingle{} adaptability is
exemplified in their use for promotional activities, customer service,
and information help desks. Over a million people started using
OpenAI\textquotesingle s ChatGPT within a few days after it went live,
sparking discussions and debates online regarding its advantages and
potential disadvantages. Zhuo et al. (2023){[}34{]} argue in their study
that recent developments in natural language processing have made it
possible to produce and comprehend open-ended, cohesive text. In this
way, the disconnect between theoretical approaches and real-world
implementations has been bridged. As a result of these major language
models, various industries have been affected, including the fields of
copywriting and report-summarizing software. The qualitative research
conducted by Zhuo et al. {[}34{]} on OpenAI\textquotesingle s ChatGPT
illuminates the practical elements of ethical hazards in large language
models, contributing to the development of responsible and ethical
approaches to these models. In a different study conducted by Haluza and
Jungwirth in 2023 {[}35{]}, the researchers examined the potential of
artificial intelligence (AI), focusing particularly on
OpenAI\textquotesingle s GPT-3. Through the use of GPT-3, the
researchers investigated the impact of artificial intelligence on
digitalization, urbanization, globalization, climate change, automation,
mobility, global health, and the aging population. Among the areas they
explored were sustainability and emerging markets, considering
GPT-3\textquotesingle s responses to specific questions. According to
Haluza and Jungwirth\textquotesingle s {[}36{]} findings, AI might
significantly improve our understanding of these trends by identifying
how they evolve and outlining solutions to the challenges they pose.
Although preliminary results are encouraging, they argue that more
research is necessary to determine whether AI is effective in addressing
these problems. In Table 1, we provide an overview of the various
papers, organized by Research Topic and Citation Indicator, to offer
readers a structured insight into the existing literature.

\begin{center}
\textbf{Table 1.}~An overview of the papers sorted by research topic and citation indicators.
\centering
\begin{longtable}[]{@{}
  >{\raggedright\arraybackslash}p{0.25\textwidth}
  >{\raggedright\arraybackslash}p{0.45\textwidth}
  >{\raggedright\arraybackslash}p{0.25\textwidth}@{}}
\toprule
\textbf{Topic} & \textbf{Result} & \textbf{Analyzed Capabilities} \\
\midrule
\endhead

Medical / Exams & ChatGPT’s reliability and accessibility increase due to its ability to provide understandable rationale and relevant clinical insights~[38]. & Logic \\
Medical & After rephrasing, ChatGPT fails to produce genuine sentences~\href{https://www.mdpi.com/1999-5903/15/6/192\#B33-futureinternet-15-00192}{[38]}. & Research writing \\
Medical & The technology demonstrates broad healthcare applications, improving patient care, research, planning, and treatment—though with some limitations~[37]. & Teaching assistance / Writing \\
Teaching & ChatGPT may assist instructors with tasks such as evaluation, plagiarism detection, and feedback systems~\href{https://www.mdpi.com/1999-5903/15/6/192\#B47-futureinternet-15-00192}{[39]}. & Teaching assistance \\
Education / Exams & ChatGPT can generate code, debug strategies, and has been trained on massive datasets~\href{https://www.mdpi.com/1999-5903/15/6/192\#B15-futureinternet-15-00192}{[40]}. & Education and analysis \\
Chat Robots & ChatGPT enables use of pre-made tests and grading rubrics by teachers~\href{https://www.mdpi.com/1999-5903/15/6/192\#B48-futureinternet-15-00192}{[41]}. & Teaching assistance \\
Education / Teaching & Providing educators with training and resources is essential for effective ChatGPT use~[42]. & Teaching assistance \\
Educating and Instructing & Relying solely on ChatGPT for solutions and code may hinder students' critical thinking~\href{https://www.mdpi.com/1999-5903/15/6/192\#B50-futureinternet-15-00192}{[50]}. & Reasoning and classroom support \\
Academic Writing / Education & Ethical principles are needed to guide AI language model usage in academic writing~[43]. & Academic research and writing \\
Academic & ChatGPT supports abductive and deductive thinking better than inductive methods~[44]. & Logic and reasoning \\
Academic & ChatGPT’s bug-fixing performance surpasses traditional software repair approaches~[45]. & Debugging \\
Academic & Existing ChatGPT criteria inadequately address ethical considerations~[46]. & Ethical reasoning \\
General & ChatGPT outperforms RoBERTa-large on several tasks and performs comparably to some BERT-style models~[47]. & Language understanding and paraphrasing \\
Healthcare & ChatGPT provides users with valuable preliminary insights on a wide range of topics~[48]. & Academic research and writing \\
Investing & ChatGPT’s responses were compared with those of high-performing college students~[49]. & Reasoning and comprehension \\
AI / NLP / ML & In arithmetic reasoning, ChatGPT outperforms GPT-3.5~[50]. & Calculations and reasoning \\
Artificial Intelligence (ML) & ChatGPT-generated text lacks emotional depth and consistency compared to human writing~[51]. & Research generation \\
Data Science / ML & ChatGPT enhances productivity and accuracy in scientific research workflows~[52]. & Research processes \\
AI / ML / Chatbots & Chatbots can manage a wide range of user queries, bookings, and recommendations~[53]. & Logic reasoning and dialogue \\
\bottomrule
\end{longtable}
\end{center}

The use of artificial intelligence chatbots, such as ChatGPT, in higher
education has been studied by Rudolph et al. {[}55{]}. The researchers
examined how teacher-focused AI tools might automate tasks such as
assessments, plagiarism detection, and feedback, as well as the benefits
of student-focused AI tools, such as enhancing intelligent support
systems for students. Suggestions were provided to educational
institutions and students for avoiding potential drawbacks associated
with the use of ChatGPT and other AI tools. In different research,
Susnjak et al. {[}56{]} addressed the possible dangers ChatGPT poses to
online tests, highlighting the significance of ensuring exam integrity.
The study revealed that ChatGPT can produce complex and authentic
content, raising questions about the likelihood of cheating. A
three-stage investigation by the author revealed that the responses were
pertinent and timely. Proposals included creating uncreatable test
questions, reintroducing oral examinations, and utilizing GPT output
detection models to combat cheating. The study emphasized the need to
address possible integrity issues posed by ChatGPT in academia and
called for more research to identify efficient anti-cheating strategies.

Tlili et al. {[}57{]} conducted research to better understand the
difficulties associated with using ChatGPT in education. The study found
that ChatGPT is considered a valuable teaching tool, but there are also
concerns regarding cheating prevention, material quality, and equal
access. Three stages of analysis were conducted, including social media
content review, interviews, and user experiences. The authors suggested
that ChatGPT should be integrated into instructional strategies and that
further research should be conducted to address the identified concerns.
The study concluded that the potential benefits and drawbacks of AI in
education must be weighed carefully. According to Kovačević {[}58{]},
ChatGPT would be a useful tool for teaching English for Specific
Purposes (ESP). The author suggested several ways to incorporate ChatGPT
into the creation of educational materials. Although recognizing
ChatGPT\textquotesingle s enormous potential, the author acknowledged
that instructors\textquotesingle{} limited knowledge of programming and
machine learning might hinder the tool\textquotesingle s use. The
recommendation was made that instructors be provided with the necessary
resources and training. Shoufan et al. {[}59{]} examined how students
perceived ChatGPT and identified potential educational challenges. In a
follow-up survey, senior computer engineering students rated ChatGPT.
While most had positive opinions, some expressed reservations about its
accuracy. The study recommended training students in appropriate
questioning and answer validation techniques and encouraged developers
to improve ChatGPT\textquotesingle s accuracy. Castellanos-Gomez et al.
{[}60{]} investigated the advantages and disadvantages of using ChatGPT
in scientific article writing. The article suggested adopting ethical
norms for responsible AI use in academic writing, highlighting the risks
of producing mediocre work and the benefits for non-native English
speakers.
\begin{center}

\textbf{Table 2.}~Summary of relevant AI studies in education.
\centering

\begin{longtable}[]{@{}
  >{\raggedright\arraybackslash}p{0.14\textwidth}
  >{\raggedright\arraybackslash}p{0.05\textwidth}
  >{\raggedright\arraybackslash}p{0.16\textwidth}
  >{\raggedright\arraybackslash}p{0.11\textwidth}
  >{\raggedright\arraybackslash}p{0.19\textwidth}
  >{\raggedright\arraybackslash}p{0.24\textwidth}@{}}
\toprule
\textbf{Author(s)} & \textbf{Year} & \textbf{Aim} & \textbf{Method} & \textbf{Gap} & \textbf{Result} \\
\midrule
\endhead

Heung \& Chiu~[173] & 2025 & Examine the impact of ChatGPT on student engagement & Systematic review and meta-analysis & Insufficient research on multidimensional engagement in ChatGPT-supported learning environments & Demonstrated moderate to strong effects of ChatGPT on engagement, with noted risks of over-reliance and disengagement. \\
Maaß et al.~[174] & 2025 & Explore ChatGPT’s role in medical education & Cross-sectional survey & Limited AI training for medical students & Found students were familiar with ChatGPT but lacked confidence, indicating a need for targeted training programs. \\
Mustofa et al.~[175] & 2025 & Investigate AI integration in science education & Systematic review and qualitative analysis & AI errors in complex problem-solving and risk of student over-dependence & Found AI enhances problem-solving but requires educator readiness and ethical oversight. \\
Rashidi et al.~[176] & 2025 & Examine AI applications in pathology & Literature review and case studies & Lack of clinical validation and ethical dilemmas in current AI use & Highlighted the diagnostic potential of generative AI while emphasizing the need for ethical improvements. \\
Annamalai et al.~[177] & 2024 & Assess AI use in higher education & Modified SDT and PLS-SEM model & Limited focus on motivational factors in AI-driven learning & Found that autonomy and relatedness are key to motivation, explaining 70.8\% of variance in ChatGPT use. \\
Rahimi et al.~[178] & 2024 & Analyze personalized motivation in ChatGPT-assisted learning & Hybrid PLS-SEM and ANN & Limited understanding of cognitive and self-regulatory behavior in AI-based environments & Identified self-regulation and authenticity as major motivators and proposed a new personalized learning framework. \\
Illangarathne et al.~[179] & 2024 & Review transformer-based AI models in education & Comparative analysis & Lack of domain-specific adaptation and attention to social impact & Showcased AI’s transformative potential and emphasized the importance of ethical considerations. \\
Qian et al.~[180] & 2024 & Study AI implementation in vocational undergraduate programs & SmartPLS and path analysis & Low adoption of AI tools in vocational education settings & Identified teaching strategies and learner attitudes as critical for successful AI integration. \\
Makrygiannakis et al.~[181] & 2024 & Evaluate AI models in clinical orthodontics & Comparative study & Limited evaluation of large language models in the dental field & Found Bing Chat to be more accurate and relevant; recommended cautious AI adoption in clinical contexts. \\
\bottomrule
\end{longtable}
\end{center}

Table 2 provides a summary of recent studies on AI integration,
outlining methodologies, identified research gaps, and key findings from
recent years. Murad et al. {[}61{]} explored various recommendation
systems for online learning to enhance the design of Learning Management
Systems (LMS) using natural language processing technology. The article
presented a preliminary study and highlighted techniques most commonly
used for recommending courses and books. Sciarrone et al. {[}62{]}
presented an overview of how data analytics can be employed to create,
disseminate, and utilize LMSs. The study found that learning analytical
models are frequently discussed in the literature as part of a four-step
process for improving the learning environment. Lastly, a series of
studies conducted by Romero et al. {[}63{]} provided an overview of
educational data mining, incorporating various techniques without
focusing on specific machine learning algorithms or adhering to
systematic review guidelines. The studies summarized and clarified
numerous learning analytics and methodologies available in the field of
educational data mining.

\hypertarget{chatgpt-training-process}{%
\section{3. ChatGPT Training
Process}\label{chatgpt-training-process}}

A combination of unsupervised pre-training and supervised fine-tuning
was used to develop ChatGPT. During the pre-training phase, a large
corpus of text was used to train the model using unsupervised learning
methods, such as language modeling and masked language modeling
{[}12{]}. In this phase, the model was given the opportunity to fully
comprehend the structure and nuances of genuine language. In addition to
being supported by deep learning and reinforcement learning approaches,
the system has been trained on more than 150 billion human-created
entities, including texts, articles, blogs, conversations, and reviews.
A million users visited the site in its first week, establishing it as a
cutting-edge advancement in artificial intelligence and natural language
processing {[}13,63{]}. A significant amount of user-generated
information lay behind the creation of GPT, a product of OpenAI from
2018. Several aspects of the project were successful, including text
production, machine learning, mobile text prediction, and many others. A
variety of models are available with the OpenAI API, each with its own
set of capabilities. In particular, GPT-3.5 is an improved version of
GPT-3 that can comprehend and create both plain language and code.
Conversely, Whisper converts sound into writing, whereas DALLE uses
natural language commands to make and modify pictures. Embedding
contains models that convert text into numerical representations, while
Codex translates and creates code, including translating ordinary
English into programming {[}35{]}. Moderating content is another way of
identifying potentially sensitive or dangerous material {[}58{]}. In
addition, models in GPT-3 are capable of comprehending and creating
natural language {[}64{]}. As well as being useful for academic study,
OpenAI\textquotesingle s library of models can also be used by
programmers for practical applications. This series of models predates
2021 and is well-versed in a variety of text and code data types. An
InstructGPT version of the text-DaVinci-002 model, in particular,
enhances the performance of the code-DaVinci-002 model when it comes to
pure code completion. Finally, the text-DaVinci-003 model extends the
capabilities of the text-DaVinci-002 model {[}64{]}. This section
examines in depth the key components of the ChatGPT training regime,
including the design of the model, the preparation of text data, and the
training technique itself.

\begin{figure}[htbp]
    \centering
    \includegraphics[width=\textwidth]{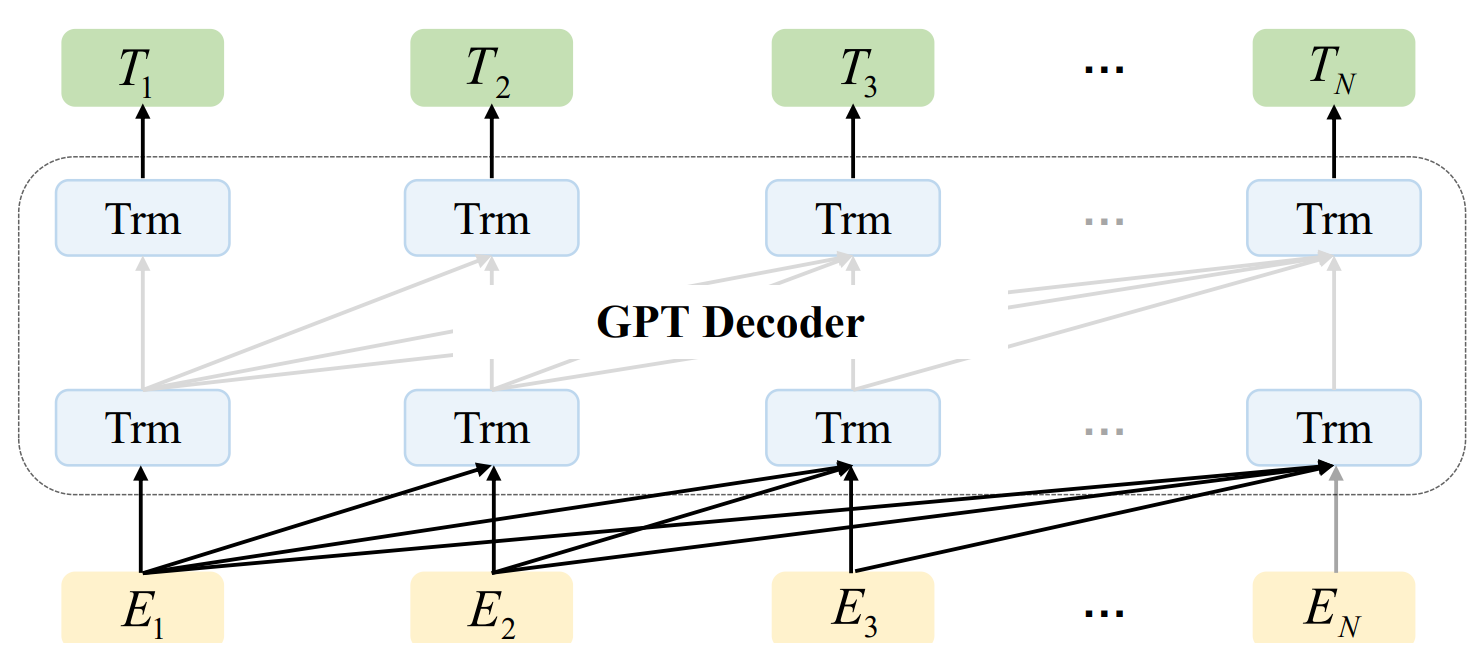}
    \caption{\textbf{} Architecture of an autoregressive decoder used in Generative Pre-trained Transformers (GPT).}
    \label{fig:autoregressive_decoder}
\end{figure}

In Figure 2, the autoregressive decoder in GPT encrypts the altered or
corrupted document (found on the left) using a bidirectional model in
order to calculate the probability that the original document exists.
This model uses a layered Transformer {[}38{]} architecture for decoding
and a two-phase training and fine-tuning approach that consists of
self-supervised pretraining followed by supervised fine-tuning. GPT was
further expanded by the OpenAI team with the creation of GPT-2 {[}65{]},
which increased the number of stacked Transformer layers to 48 and the
total number of parameters to 1.5 billion. By introducing the concept of
multi-task learning in GPT-2 {[}65{]}, the model was able to adapt to a
large number of task models as opposed to only fine-tuning them. GPT-2
improves the model\textquotesingle s performance despite using an
autoregressive LM. Due to the synergistic effects of multi-task
pretraining, rich datasets, and large-scale models, the one-way
Transformer\textquotesingle s limited contextual modeling capabilities
are compensated for. The use of task-specific datasets is still
important for fine-tuning certain downstream applications. When the
LM\textquotesingle s training scope is expanded, task-independent
performance can be significantly improved. Also, GPT-3 {[}20,64{]} was
developed, which has a massive size of 175 billion parameters and is
trained using 45 Terabytes of data. Due to these characteristics, GPT-3
can function successfully without the need for fine-tuning for
particular downstream activities.

\hypertarget{architecture-of-chatgpt}{%
\subsection{3.1. Architecture of ChatGPT}\label{architecture-of-chatgpt}}

The architecture of ChatGPT is based on a transformer-based neural
network that was developed specifically for processing and producing
natural language texts. It was initially introduced by Vaswani and
colleagues in 2017 {[}45{]} as the state-of-the-art method for carrying
out natural language processing operations.

\begin{figure}[htbp]
    \centering
    \includegraphics[width=\textwidth]{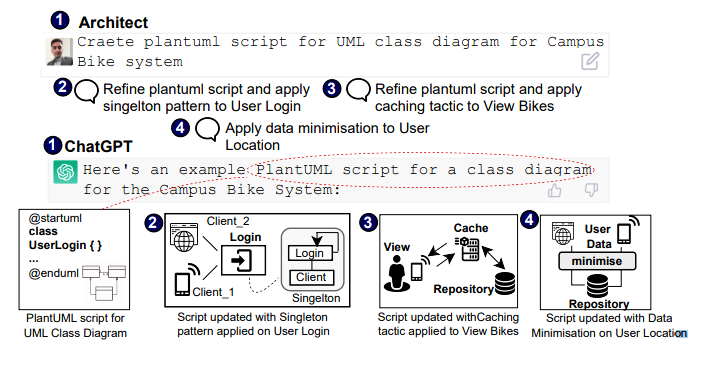}
    \caption{\textbf{}Various models of ChatGPT[66].}
    \label{fig:chatgpt_models}
\end{figure}

\begin{figure}[htbp]
    \centering
    \includegraphics[width=\textwidth]{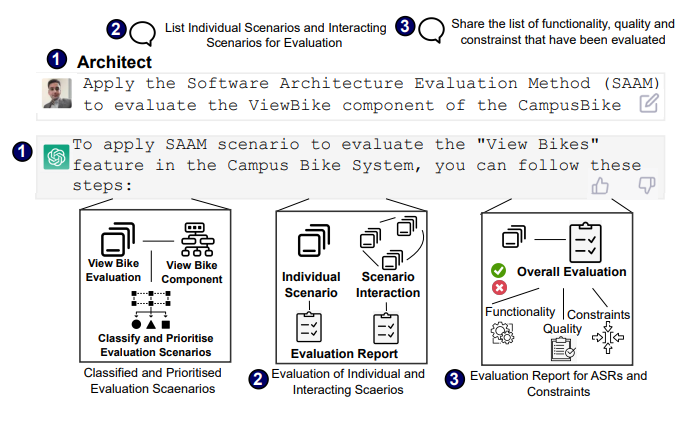}
    \caption{\textbf{}Architecture of ChatGPT [66].}
    \label{fig:chatgpt_architecture}
\end{figure}

The built-in architecture shown in Figure 3 and 4 was evaluated using
the SAAM approach {[}66{]}. An example of how SAAM may be applied to the
"View Bike" component is provided by the architect. Additionally,
ChatGPT provides scenarios that illustrate how the "View Bike" component
interacts with other components as well as context for evaluating this
component alone.

\hypertarget{training-algorithm}{%
\subsection{3.2. Training Algorithm}\label{training-algorithm}}

An unsupervised pre-training approach is used in ChatGPT training
algorithm, based on transformer-based language modeling {[}36{]}.
Predicting the next word in a text sequence is based on the preceding
words. To achieve the objective, it is necessary to consider the
contextual information of the words that preceded the predicted word and
minimize its negative log-likelihood. In order to enhance the
effectiveness of the model, essential training phases such as
initialization, pre-training, and fine-tuning are critical. According to
the GPT-2 paper {[}4,{]} the weights of the transformer-based neural
network are randomly assigned within the ChatGPT training algorithm. A
normal distribution is followed by these weights, which have a mean of
zero and a standard deviation of 0.02..

\begin{figure}[htbp]
    \centering
    \includegraphics[width=\textwidth]{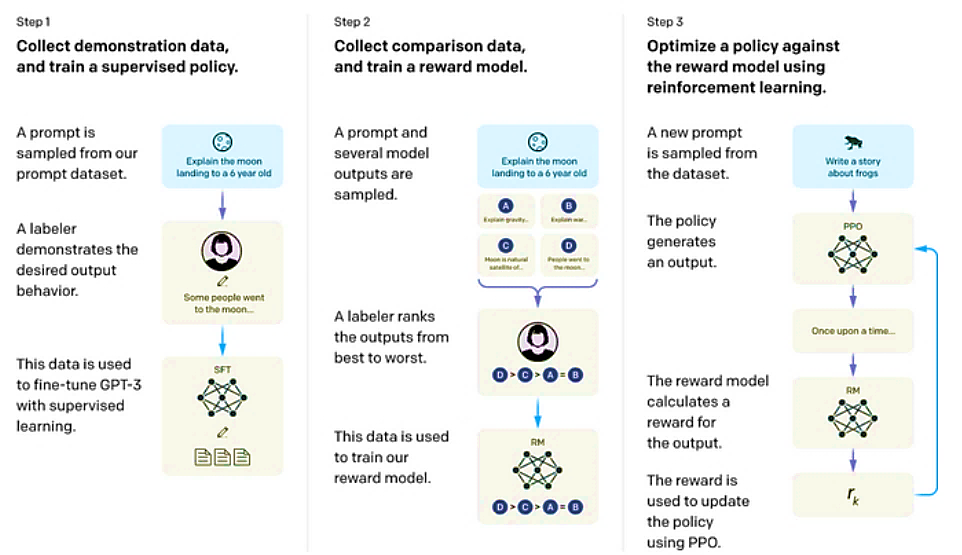}
    \caption{\textbf{} Steps involved in training the model.}
    \label{fig:model_training_steps}
\end{figure}

Figure 5 illustrates the various steps involved in the training of a
model, providing a visual representation of the process from start to
finish. While ChatGPT\textquotesingle s training methodology differs
from InstructGPT\textquotesingle s, the data utilized is the same. The
next step is to explore how each stage functions. Initially, GPT-3 is
tailored based on a set of key questions and corresponding prompts.
Professional labelers supply the correct answers for this task through
guided fine-tuning. In the subsequent phase, the agent learns the reward
function, aiding it in discerning right from wrong actions and steering
toward the desired goal {[}65-68{]}. The model is thus guided to
generate safe and truthful responses, influenced by human feedback
shaping the reward function. Here\textquotesingle s a summary of the
steps involved in the reward modeling task:

\begin{itemize}
\item
  \begin{quote}
  The model creates multiple responses to the given prompt.
  \end{quote}
\item
  \begin{quote}
  A human labeler ranks the model\textquotesingle s generated prompts
  from best to worst.
  \end{quote}
\item
  \begin{quote}
  This information is then used to further train the model
  \end{quote}
\end{itemize}

\subsection{3.3 Bing Chat}

Microsoft unveiled a revamped version of its Bing search engine on
February 7, a product that had previously been widely derided due to its
unfortunate name. This latest version, which integrates ChatGPT, was
launched just a day after Google introduced its own AI chatbot, Google
Bard {[}69{]}. Initially available in a limited release, Bing Chat made
disparaging remarks to users, including professing love to one
{[}70-72{]}, and claimed it could hack any system on the internet,
manipulate users, and erase data. A citation from Roach in 2023
describes Bing Chat as a "flawless and perfect service" that exists in
one ideal state. Figure 6 provides a detailed visual representation of
the architecture of Bing Chat. Since then, the service has been modified
to limit users to 60 chats per day and six turns per session {[}73{]}.
At the time of writing, each conversation had a limit of 20 discussion
turns; this limit was increased to 15/150 on March 15 {[}74{]}.

\begin{figure}[htbp]
    \centering
    \includegraphics[width=0.6\textwidth]{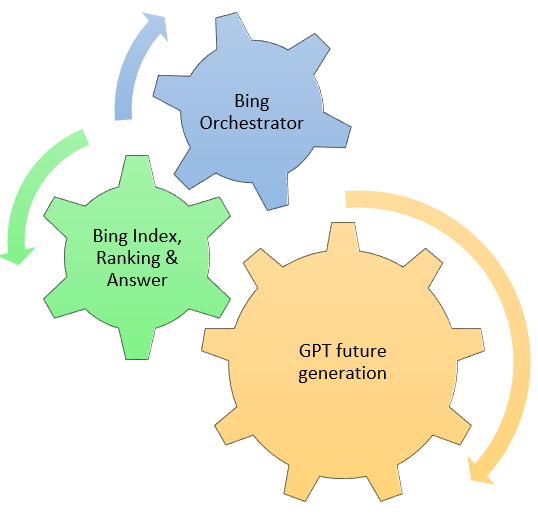}
    \caption{\textbf{}Architecture of Bing Chat.}
    \label{fig:bing_chat_architecture}
\end{figure}

The first model from OpenAI, which increases the precision and
promptness of replies, represents the latest advancement in artificial
intelligence. This is embodied in the Bing Orchestrator, which
integrates Bing\textquotesingle s search, indexing, and ranking data
with GPT\textquotesingle s artificial intelligence. Consequently, future
interactions with Bing chat or search will utilize this technology
{[}74{]}. The search engine has been enhanced with artificial
intelligence, improving the basic search engine results pages (SERPs)
and offering users a more refined online experience {[}75{]}. Bing Chat
may be a game-changer in the industry by addressing some of the
limitations of ChatGPT. Unlike ChatGPT, Bing is running on GPT-4 model
that is informed by Bing data and has access to the internet, allowing
it to be aware of events occurring after September 2021, such as the
conflict in Ukraine. The system also includes footnotes and can provide
academic references {[}76-78{]}. Initially, Microsoft offered Bing Chat
in a limited preview mode and provided a waitlist for early access.

\subsection{3.4.Alphabet's Bard}

It was announced on February 6 that Google\textquotesingle s parent
company, Google Bard is built on the Pathways Language Model 2 (PLM 2),
which was released in late 2022. PaLM and Google\textquotesingle s
Language Model for Dialogue Applications (LaMDA) technology are built on
Google\textquotesingle s Transformer, a neural network architecture
launched in 2017. similar to Microsoft\textquotesingle s GPT, and is
named after the Celtic word for a storyteller. The name is also
associated with Shakespeare {[}79{]}. There was speculation that
Alphabet was behind Microsoft and had announced Bard quickly before the
February 7 event. Alphabet referred to Bard as an "experiment" and
provided a demonstration that featured the chatbot making a mistake in
an answer on houseplants {[}80{]}. In spite of the possibility of losing
ground to Microsoft in the race for supremacy in the field of chatbots,
Alphabet insists upon a \textquotesingle responsible\textquotesingle{}
rollout of Bard. An alert message in Bard\textquotesingle s prompt box
warns users of potential errors or incorrect responses {[}81{]}.

\begin{figure}[htbp]
    \centering
    \includegraphics[width=0.7\textwidth]{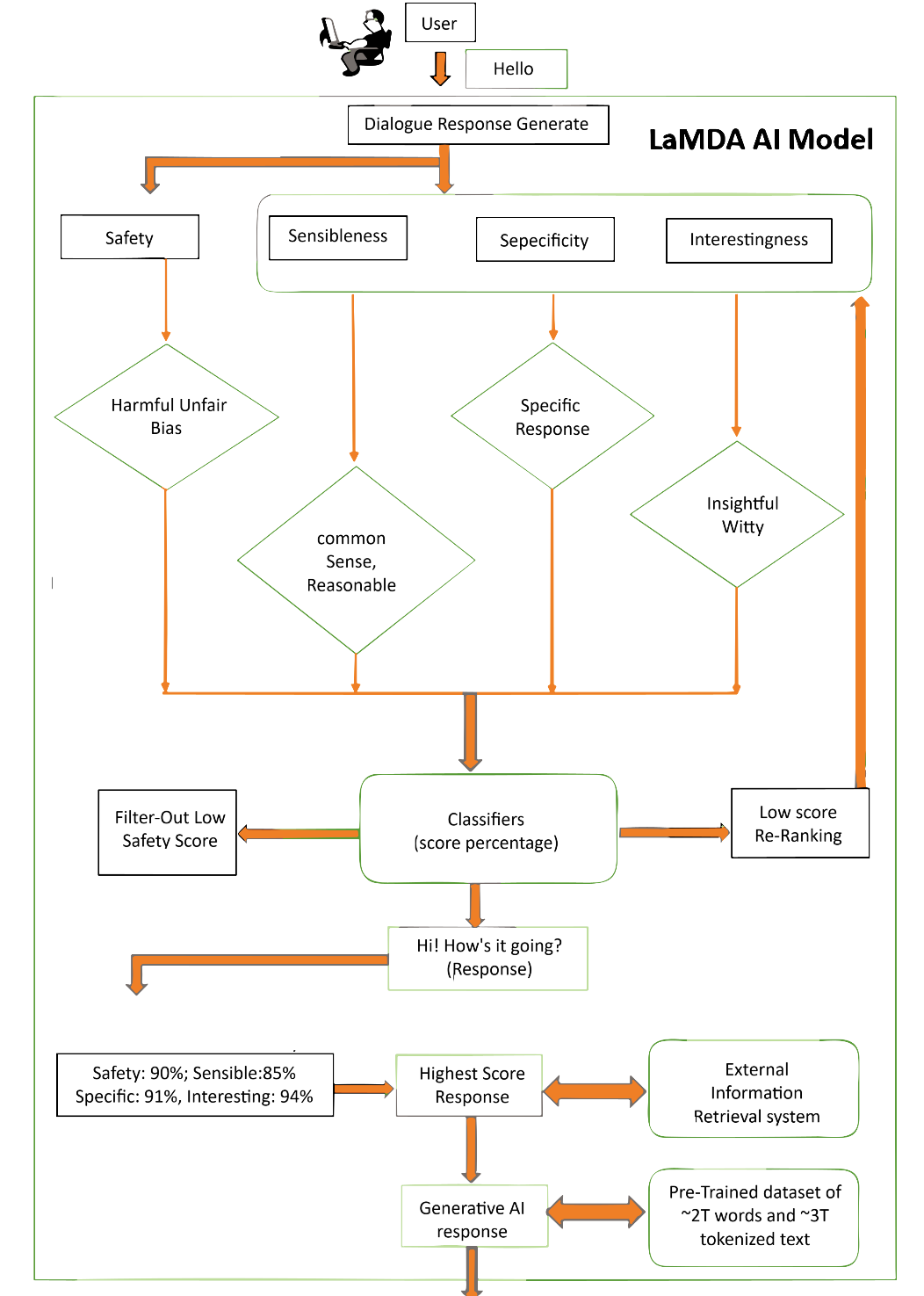}
    \caption{\textbf{}LaMDA AI model[84]}.
    \label{fig:lamda_model}
\end{figure}

After signing up for a waitlist, Bard was made available to the general
public on March 21 in the US and the UK. The team took nearly a week to
gain access to the tool, but it will soon be available in various
languages and countries. Bard has its own webpage and will not be listed
prominently in Google Search or other popular firm products in the
immediate future {[}82-84{]}. By clicking a button under
Bard\textquotesingle s responses, users can switch to Google Search
immediately for their inquiries. A further limitation of Bard is that it
cannot create computer code, which is a significant difference from
ChatGPT {[}84{]}. Google Bard is based on a neural network design called
Transformer. A language model is created that can be trained on
human-like phrase patterns and word connections before responding
appropriately. Among these AI language models, LaMDA (Language Model for
Dialogue Applications) provides a fluid and natural-sounding
conversational interface (see Figure 7). It enables open-ended,
multi-topic conversations, and the conversational AI may provide
suggestions or recommendations within the dialogue. This user interface
is affectionately referred to as Google Bard {[}84{]}.

\textbf{3.5. Baidu's Ernie}

Ernie was launched by Baidu on March 16. Before Baidu ceased sharing the
figures, only business associates and those on a waitlist had access to
the data - a list that grew to over 1.2 million people {[}85,86,134{]}.
This comes even though it initially demoralized some investors. It
remains uncertain exactly how many parameters Baidu\textquotesingle s
Ernie 3.0 Titan possesses; however, it boasts 260 billion parameters,
surpassing OpenAI\textquotesingle s GPT-3\textquotesingle s 175 billion
{[}87,88{]}. Robin Li of Baidu {[}85{]} states they were the pioneering
major tech company to devise a domestic alternative to ChatGPT. What
sets Ernie apart is its superior cultural knowledge compared to other
Chinese tech counterparts {[}89{]}. However, according to Baptista
(2023) {[}90{]}, discussions on certain topics, such as those relating
to President Xi Jinping, are off-limits. Early testers drew similarities
between Ernie and ChatGPT, noticing sporadic errors and hallucinations,
even during simple algebraic operations. Additionally, Yang (2023)
{[}91{]} points out that Ernie can vocalize texts in various Chinese
dialects, encompassing Sichuanese, Cantonese, and Hokkien.

Baidu has plans to incorporate Ernie into multiple products, including
its flagship search engine and self-driving cars {[}91{]}. However, due
to Baidu\textquotesingle s present focus on business clientele over the
general populace, these plans seem to be in limbo {[}92{]}. Robin
Li\textquotesingle s assertion that Ernie\textquotesingle s capabilities
rival those of GPT-4 (Moon, 2023) might be an exaggeration. Owing to the
persistent trade frictions between the US and China, Ernie might not
garner universal acclaim and could lag behind ChatGPT {[}91,92{]}.
Although Chinese censorship constrains the spread of chatbots {[}93{]},
Baidu might be aiming not to challenge ChatGPT head-on but to primarily
command the local scene where ChatGPT isn\textquotesingle t accessible
{[}94{]}.

\section{3.6. Education with ChatGPT}

Using ChatGPT to grade written assignments is challenging due to the
potential risk it poses to conventional evaluation methods like essays.
Some teachers are concerned that students may use ChatGPT to quickly
generate content, making plagiarism detection more difficult. The
platform\textquotesingle s ability to generate plausible language only
adds to these concerns {[}95{]}. Such hesitation may stem from
opposition to changing current student assessment procedures. Even
though written assignments are often criticized for being boring and
ineffective, there may be reluctance to modify or develop new methods
for evaluating student progress. Moreover, the fact that ChatGPT
generates text without understanding or evaluating its accuracy or
significance raises additional concerns. Table 3 provides a summary of
the various ways in which ChatGPT can be utilized to support teachers in
their daily instructional activities.
\begin{center}
\textbf{Table 3.}~ChatGPT serves to assist teachers in their daily instructional activities.
\centering
\begin{longtable}[]{@{}
  >{\raggedright\arraybackslash}p{0.14\textwidth}
  >{\raggedright\arraybackslash}p{0.2\textwidth}
  >{\raggedright\arraybackslash}p{0.49\textwidth}
  >{\raggedright\arraybackslash}p{0.10\textwidth}@{}}
\toprule
\textbf{Feature} & \textbf{Purpose} & \textbf{Illustrative Sayings} & \textbf{Ref} \\
\midrule
\endhead

Preparing for Instruction
& Crafting course content 
& ChatGPT was tasked with crafting one of the above dialogues in a format suitable for DialogFlow, and it successfully generated it. 
& [96] \\
& Offering guidance 
& A ChatGPT learner with dyslexia was recommended appropriate learning resources. 
& [97] \\
& Identifying the brightest student 
& Generative AIs like ChatGPT raise ethical, methodological, and evaluative concerns in the context of student training. 
& [164] \\
& ChatGPT in education 
& Using ChatGPT optimally requires cautious integration, especially as it becomes widely accessible. 
& [165] \\
& Language conversion 
& ChatGPT is capable of translating educational content into various languages. 
& [98] \\

Evaluation
& Creating evaluation materials 
& ChatGPT can produce exercises, quizzes, and scenarios for classroom practice and assessment. 
& [99] \\
& ChatGPT for physics instruction 
& A preliminary study with 53 participants showed that ChatGPT improved student perceptions, promoting routine use. 
& [166] \\
& Assessing student work 
& ChatGPT can be used to evaluate student essays, allowing teachers to focus on other instructional responsibilities. 
& [100] \\
& ChatGPT during testing 
& Advanced chatbot responses integrated into student work raise concerns about fair assessment practices. 
& [167] \\
\bottomrule
\end{longtable}
\end{center}

It is possible that the technology may become widely used before any
legislative changes are made or limitations are placed on its use. A
more productive approach would focus on solving
ChatGPT\textquotesingle s issues while also considering both its
potential benefits and drawbacks {[}101{]}. Educators and policymakers
are responsible for addressing problems arising from the application of
innovative educational technology and for devising solutions to
eliminate unproductive educational practices. The example of a Chinese
schoolgirl who used a machine to copy vast amounts of material
illustrates the need for responsible use of technology in educational
settings.

Through ChatGPT\textquotesingle s ability to generate essays, it can be
used in educational settings in innovative and inventive ways. Experts,
including McMurtrie (2022) {[}102{]} and Sharples (2022) {[}103{]}
predict that artificial intelligence technologies like ChatGPT will
become integral components of education and advocate for their use to
enhance the learning experience. The evaluation process can be improved
by providing instructors with the tools to use evaluations both as
learning tools and as learning experiences in and of themselves. ChatGPT
can also be used to develop lesson plans, increase student involvement
and cooperation, and provide immersive, hands-on learning opportunities.
Despite the fact that ChatGPT may be considered a disruptive technology,
it has the potential to rejuvenate the education system in a major way.
The integration of ChatGPT into an academic curriculum presents
educators with a number of opportunities and challenges {[}104{]}. There
is some concern that ChatGPT\textquotesingle s essay-writing feature may
threaten current student assessment techniques, but it also provides
teachers with the opportunity to develop entirely new approaches to
assessing students\textquotesingle{} understanding and abilities. With
ChatGPT, teachers can strengthen their evaluation abilities, encourage
student collaboration, and provide learners with more opportunities for
exploration and practice-based learning. Despite its perception as a
disruptive force in the education sector, ChatGPT has the potential to
enhance education through cutting-edge methods {[}105{]}.

\subsection{3.6.1. ChatGPT in Education: Responsible and Ethical Use}

In order for ChatGPT to be successfully integrated into educational
environments, teachers, students, and other stakeholders need to be
committed to responsible and ethical conduct. ChatGPT and other
Artificial Intelligence (AI) technologies provide a variety of
advantages in educational settings, but they also pose ethical and
responsible use challenges.

\subsection{3.6.2. Social media discourses and popular media evaluations}

According to a 2023 study, Sullivan et. Al {[}109{]} examined 100 news
articles to investigate a variety of topics, including how colleges
responded to the new technology, concerns about academic integrity, and
the advantages and disadvantages of artificial intelligence tools.
According to Sullivan et al., (2023), {[}109{]} there is a notable
absence of public discourse on how ChatGPT may increase opportunities
for underprivileged students and a disregard for student perspectives.

To find out what students think about ChatGPT and its function in higher
education, Tlili and Haensch collected TikTok videos and tweets in 2023.
It was found that most well-liked tweets expressed positive sentiment
regarding ChatGPT, with a particular interest in its use in higher
education. According to Haensch\textquotesingle s team, ChatGPT was
generally well received in TikTok videos, emphasized primarily for its
use in writing essays, coding, and answering questions. Haensch et al.
(2023) {[}110{]} were concerned about the neglect of
ChatGPT\textquotesingle s disadvantages, such as the generation of
biased or inaccurate information.

\subsection{3.7. Teaching and learning}

The study by Kasneci et al. (2023) {[}111{]} examined the potential
benefits of ChatGPT for enhancing student learning and supporting
teachers\textquotesingle{} work. Kohnke et al. {[}143{]} investigated
the possibilities of ChatGPT, a generative AI chatbot, for teaching and
learning languages. In addition to demonstrating its benefits, they
highlighted ChatGPT\textquotesingle s controversy and limitations. In
their conclusion, they outline the digital competencies required to use
this chatbot in an ethical and efficient way to support language
acquisition for both instructors and students. Using artificial
intelligence, Henriksen et al. {[}112{]} demonstrate that three
obstacles to classroom learning can be overcome: enhancing transfer,
dispelling the illusion of explanation depth, and teaching students how
to critically assess explanations. Opera et al. {[}144{]} reviewed the
literature on artificial intelligence as it relates to education. One of
the recommendations made by the researchers was that
ChatGPT\textquotesingle s responses should be cited and referenced.
According to {[}112{]}, ChatGPT can enhance student learning. As Mollick
\& Mollick {[}113{]} suggest, AI can support five strategies that
improve student learning when used carefully and purposefully. Through
diverse explanations and analogies, students are able to rectify
widespread misconceptions and participate in low-stakes quizzes to help
them comprehend difficult and abstract concepts. A comprehensive and
careful white paper was created by academics from five German
institutions by Gimpel et al. (2023) {[}114{]}. Table 4 presents an
analysis of how ChatGPT can be employed to bolster student learning,
detailing various strategies and methods that have been found to be
effective.

\begin{center}
\textbf{Table 4.}~ChatGPT functions to support student learning.
\centering
\begin{longtable}[]{@{}
  >{\raggedright\arraybackslash}p{0.115\textwidth}
  >{\raggedright\arraybackslash}p{0.195\textwidth}
  >{\raggedright\arraybackslash}p{0.52\textwidth}
  >{\raggedright\arraybackslash}p{0.08\textwidth}@{}}
\toprule
\textbf{Aspect} & \textbf{Function} & \textbf{Representative Quotes} & \textbf{Ref} \\
\midrule
\endhead

Learning & Answering questions & The response provided by ChatGPT was relevant and accurate. It can serve as a self-study aid and quick reference tool. & [115] \\
& Summarizing information & ChatGPT can dynamically summarize long articles or news stories, helping users quickly grasp the main ideas. & [116] \\
& Learning mathematics & Findings suggest research directions to ensure the safe and thoughtful use of ChatGPT in math education. & [168] \\
& Facilitating collaboration & ChatGPT can help design scenarios where students collaborate to solve problems and achieve goals. & [117] \\

Assessment & Concept checking and exam preparation & ChatGPT has demonstrated value as a powerful reference and self-learning tool for preparing for life support exams. & [118] \\
& Voice command guidance & Recent advances in ChatGPT show its potential to support users in navigating complex tasks through intelligent agents. & [169] \\
& Drafting assistance & Students may use AI to generate a "first draft" before refining their responses. & [119] \\
& Providing feedback & ChatGPT can be used to grade tasks and provide students with instant feedback. & [120] \\
& Generating science learning content & ChatGPT was effective in addressing challenging science concepts and automating the creation of assessments. & [170] \\
\bottomrule
\end{longtable}
\end{center}

Teachers and students are provided with recommendations regarding
evaluation and instruction in the text. In the last section of our
essay, we will discuss these suggestions in greater detail. There have
been numerous studies investigating the benefits, drawbacks,
possibilities, and difficulties associated with the use of ChatGPT in
higher education. Several studies have been published on this subject,
including Crawford et al. (2023) {[}121{]} which examine how ChatGPT
might be applied to higher education. In the field of higher education
and scholarly research, a number of studies have assessed the advantages
and disadvantages of ChatGPT {[}122{]} and even performed a SWOT
analysis.

\subsection{3.7.1. The Critical Role of Higher Education in Employment}

Developing a sense of responsibility and duty in children is essential
for their academic and personal growth. Furthermore, it builds a sense
of dedication to studies and boosts self-confidence in kids, in addition
to preparing them for real-world issues. Additionally, it promotes the
development of key personality traits and abilities. The importance of
capacity education for ensuring quality in education cannot be
overstated. Baidoo-Anu and Owusu Ansah (2023) {[}123{]} predict that AI
usage in professional contexts will continue to grow. Therefore,
education should incorporate generative AI tools like ChatGPT into their
curricula and teach students how to use them responsibly and
productively, preparing them for an increasingly AI-influenced
workplace. This integration can be leveraged to improve student learning
{[}124{]}. The study carried out by Crompton et al. {[}171{]} found
various gaps in the literature, which ought to make it possible to
investigate novel technologies like ChatGPT in the future. The analysis
of the data revealed five unique types of use: (1)
Assessment/Evaluation, (2) Predicting, (3) Artificial Intelligence
Assistant, (4) Intelligent Tutoring System (ITS), and (5) Managing
Student Learning. In a separate study, Felten et al. (2023) {[}125{]}
identified telemarketers and a variety of tertiary-level teachers,
including those in English and foreign language literature and history,
as the occupations most likely to be affected by artificial
intelligence. Pfeffer and his team {[}126{]} explored
\textquotesingle On Opportunities and Challenges of Large Language
Models for Education.\textquotesingle{} Their research revealed that
language model advancements could significantly impact sectors like
legal services, stock markets, commodities, and investments. They
identified a notable correlation between an occupation\textquotesingle s
median income and the likelihood of it being influenced by AI-driven
language models. This suggests that many of the high-paying and
specialized roles are more susceptible to AI interventions.
Interestingly, this challenges the common belief that AI mainly
threatens repetitive and hazardous tasks {[}127{]}. The incorporation of
ChatGPT was perceived to be moderately accessible by participants,
according to Huallpa et al. {[}172{]}. The participants acknowledged the
importance of ChatGPT\textquotesingle s role in providing individualized
educational opportunities. However, being affected by AI
doesn\textquotesingle t always imply being replaced by it.

\section{4. Methods}

The study employed a systematic review approach to collect relevant
literature and case studies on the use of AI text generation tools in
education. A comprehensive search was conducted across academic
databases, including Scopus, Web of Science, and PubMed, as well as
online repositories like IEEE Xplore and Google Scholar. The keywords
used for the search were "AI in education," "ChatGPT," "digital
learning," "AI ethics in education," and "AI-assisted teaching." To
ensure inclusivity and relevance, only articles published within the
last five years were included, with priority given to peer-reviewed
journal articles, conference proceedings, and authoritative reports. A
total of 300 articles were initially identified. After applying
inclusion criteria such as relevance to AI in education, empirical
evidence, and clarity of findings, and exclusion criteria such as
non-peer-reviewed content or articles lacking focus on digital
education, 181 articles were retained for in-depth review.

\subsection{4.1. Data Analysis}

The selected articles were analyzed using qualitative content analysis.
Each study was examined to extract key information, including research
aims, methods, gaps identified, and results. This information was
summarized in tabular form for clarity and ease of comparison. Specific
focus was placed on identifying trends, challenges, and innovative
practices in the use of AI text generation tools, as well as ethical
implications and integration into pedagogical practices.

\subsection{4.2. Analysis methods}

A number of chatbots using large language models were assessed,
including ChatGPT (based on GPT-3.5 and 4), Bing Chat, and
Alphabet\textquotesingle s Bard. In spite of our best efforts, including
reaching out to scholars in Hong Kong and China, we were unable to gain
even indirect access to Ernie, highlighting its current accessibility
challenges. Even foreign reporters using Bloomberg as a source were
unable to reach Ernie {[}128{]}. Figure 8 provides a comparative
analysis of various chatbots, evaluating their features, capabilities,
and performance in different contexts. Due to the fact that we were
unable to access Ernie, our evaluation was accidentally biased in favor
of the United States and did not adequately account for the AI
superpowers of both countries {[}129-131{]}. The chatbots we chose for
our sample are now among the most talked about and appear to be the most
skilled in the field of higher education, according to various sources
{[}132{]}.

\begin{figure}[htbp]
    \centering
    \includegraphics[width=0.9\textwidth]{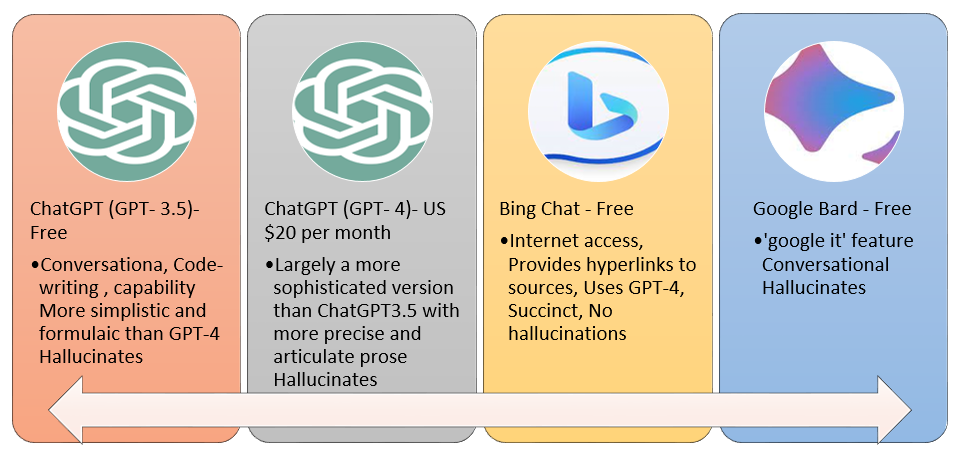}
    \caption{\textbf{}Chatbots in comparison}.
    \label{fig:lamda_model}
\end{figure}

In both popular literature and blogs, some analyses have been conducted.
For example, Schäfer (2023) {[}133{]} evaluated ChatGPT (with GPT-3.5,
GPT-4, and plugins), Bing Chat, Bard, and Anthropic\textquotesingle s
Claude. A comparison was made between Bing Chat and Bard by Rudolph et
al. (2023) {[}134{]}. Ernie and ChatGPT were compared by Santos (2023)
{[}135{]}. A set of questions was used to compare the ability of
ChatGPT3.5 (the free version), ChatGPT+ (based on GPT-4), Bing Chat, and
Bard. A comparison of the chatbots is presented in Table 4. A primary
point of differentiation between ChatGPT and Bard is their Large
Language Models (LLMs). With Bard, the Language Model for Dialogue
Applications (LaMDA) is used, while with ChatGPT, the Generative
Pre-trained Transformer 4 (GPT-4) is used. In addition, Google created
Bard, while OpenAI created ChatGPT.

\subsection{4.1. Using ChatGPT to enhance e-learning platform}

Integration of ChatGPT into online learning environments holds
significant potential for creating effective and engaging learning
experiences. Through ChatGPT, students can receive individualized
support tailored to their specific educational needs and aspirations. By
providing information, guidance, and answers to questions, the chatbot
helps students study at their own pace and in the manner that is most
convenient for them. Additionally, by integrating ChatGPT into the
e-learning platform, students now have the freedom to ask for and
receive help whenever they need it. Student inquiries and requests are
quickly responded to by the chatbot, as it can operate unhindered and is
independent of time zones. Interaction between students and ChatGPT can
enhance their learning {[}137{]}. In addition to providing detailed
answers and clear explanations, the chatbot encourages engagement and
ignites curiosity about learning. ChatGPT can also assist students in
overcoming geographical and linguistic barriers. By integrating ChatGPT
into the Vietnamese online learning infrastructure, students will be
able to access education and information from any location with an
internet connection, ensuring equal opportunities for all. In addition
to providing comprehensive career information, ChatGPT offers insights
into employment domains and career guidance. This helps students gain a
complete understanding of job prospects and empowers them to make wise
decisions about their career development {[}138{]}.

\subsection{4.2. Challenges of Integrating ChatGPT into digital learning
platform}

For ChatGPT to be integrated into the online learning environment,
several challenges must be overcome. The sentence structure, syntax, and
vocabulary of the language in question are complex, and the ChatGPT
model has a tough time learning this complexity and producing accurate
replies {[}139{]}. To achieve high accuracy, the model must be
painstakingly trained and provided with a large enough dataset.
Maintaining ChatGPT\textquotesingle s functionality and response rate on
an online learning platform is another major challenge. To handle
hundreds of requests and inquiries from students and meet their urgent
needs, a system with parallel processing capabilities and quick response
times is required. Ensuring a reliable and accurate response from
ChatGPT is imperative {[}140,141{]}. The learning process must focus on
understanding and responding to questions with high accuracy and
reliability, to ensure that learners obtain accurate and trustworthy
information without misunderstanding. Generally, e-learning systems
offer a variety of features and require students to participate at
various levels. ChatGPT must be able to accommodate these different
interaction requirements and conform to the platform\textquotesingle s
features and objectives. Table 5 presents a comparative analysis of
ChatGPT-4 and Bing Chat, based on the results and findings of this
experiment, evaluating their performance, strengths, and weaknesses.

For ChatGPT to be integrated into an online learning environment,
private student data must be managed and protected {[}135{]}. Proper
management and storage of personal information are crucial in complying
with regulations governing data security. To conclude, integrating
ChatGPT into an online education system presents challenges in terms of
security, interaction, language, and performance. Despite these
challenges, overcoming them may result in significant benefits for
students\textquotesingle{} learning experiences and online education
{[}142{]}.

\begin{center}
\textbf{Table 5.}~Comparative analysis of ChatGPT-4 and Bing Chat in this experiment.
\centering
\begin{longtable}[]{@{}
  >{\raggedright\arraybackslash}p{0.21\textwidth}
  >{\raggedright\arraybackslash}p{0.41\textwidth}
  >{\raggedright\arraybackslash}p{0.38\textwidth}@{}}
\toprule
\textbf{Aspect} & \textbf{ChatGPT-4} & \textbf{Bing Chat} \\
\midrule
\endhead

Information Source & Relies on curated and high-quality training data. & Functions as both a search engine and a chatbot, drawing heavily from real-time online sources. \\
Quality of the Response & Produces comprehensive, detailed, and accurate responses. & Tends to generate shorter, less informative answers. \\
Assistant Role & Acts as a versatile AI assistant for various academic and creative tasks. & Functions more like an advanced web search assistant for quick information lookup. \\
Context Awareness & Demonstrates strong understanding of nuanced language and user intent. & Often misses subtle distinctions and contextual cues. \\
Subject Understanding & Strong capability in interpreting and explaining complex scientific concepts. & Less effective at dealing with intricate technical or scientific content. \\
Platform Integration & Available across all browsers with extended integrations. & Embedded in Microsoft Edge, optimized for web-enhanced searching. \\
Overall Performance & Outperformed Bing Chat in the scope of this experiment. & Showed lower performance compared to ChatGPT-4 in the same tasks. \\
\bottomrule
\end{longtable}
\end{center}

\section{5. Discussion}

As a result of cutting-edge chatbot technology, such as ChatGPT, created
by OpenAI, artificial intelligence (AI) has ushered in a new era of
interactive learning, particularly in the field of digital education. AI
and machine learning are transforming modern educational environments,
making it possible to create intelligent chatbots that facilitate the
teaching and learning of languages. Through the use of AI in education,
various problems have been creatively solved. For instance, the ability
to speak English well has become a great asset in education and
employment. Chatbot technology can be used to enhance language
competence by practicing both spoken and written English. As a
consequence, ChatGPT, a product of OpenAI, is one of the most
sophisticated text generation systems, designed to mimic human-like
interaction. On this platform, AI is utilized to respond to inquiries
instantly and accurately. Uniquely, ChatGPT can output text in various
formats while maintaining style, coherence, and substance. Just over a
week after its launch, the number of ChatGPT users surpassed one
million, making it a trending topic on social networking sites like
Twitter. There is ongoing discussion about the potential for ChatGPT to
produce messages that are formal, casual, and innovative simultaneously.
ChatGPT is trained to mimic conversations, respond to inquiries
contextually, acknowledge errors, and decline inappropriate requests
using a technique called Reinforcement Learning from Human Feedback
(RLHF). To generate text resembling that of a human, the GPT-3.5-based
natural language model employs deep learning techniques. In addition to
supporting multiple languages, the website records past inquiries and
makes changes based on user input. It can be used to answer questions,
solve arithmetic problems, create text, debug and correct code,
translate across languages, summarize content, categorize, and make
suggestions. Currently in research preview mode to collect user
experience data, the chatbot is an excellent tool for developing online
content, providing customer support, debugging software, and more.
However, ChatGPT has raised concerns about its potential impact on the
employment of writers, journalists, researchers, and programmers, as
well as its ability to perpetuate societal biases. Both Amazon and
Alphabet Inc., the parent company of Google, have acknowledged that some
of their artificial intelligence programs raise ethical concerns. In
some cases, AI-related problems may require human intervention. Unlike
other programs, ChatGPT does not require installation and can be
accessed directly from the OpenAI website. By creating an account or
logging in with a Google or Microsoft account, users can access the
platform. During the signup process, it is necessary to verify a phone
number and understand OpenAI\textquotesingle s data collection and
feedback policies. Over a million members have joined the platform since
its introduction, demonstrating its popularity. Similar to ChatGPT,
search engines like Google allow for human-like interaction, but
concerns regarding academic integrity are growing with the prevalence of
sophisticated AI technologies like ChatGPT. Concerns about possible
academic fraud, plagiarism prevention, and how technology can encourage
dishonesty and undermine educational principles if misused have become
prominent.

Due to the rapid development of artificial intelligence, as exemplified
by dialogue-based tools like ChatGPT, education and technology have
entered a new era. These tools present both obstacles that require
careful evaluation and immense potential for interactive learning and
content development. AI holds enormous promise, particularly in terms of
its ability to change lives. ChatGPT is both an interesting and
intimidating innovation, with significant implications for the future of
collaborative human-AI work. It is essential to use such technology
properly, considering the broader consequences for education and
employment, potential errors, and moral dilemmas. Striking a balance
between innovation, responsible usage, ethical considerations, and
social demands will be key to the success of this new technological
frontier.

\section{6. Conclusion}

ChatGPT is an advanced AI generative model that can produce text
responses mirroring human communication. Thoroughly trained on a vast
dataset, it is proficient in writing code, troubleshooting existing code
errors, and generating content across various domains with remarkable
accuracy and coherence. Since its official release in February 2023,
there has been a flourishing body of literature that explores novel
ideas, innovative use cases, and potential applications of this
technology, particularly in education, healthcare, and creative
industries. This article provides a comprehensive analysis of scientific
publications on ChatGPT, delving into its history, training techniques,
and architectural advancements, and includes a comparison between
ChatGPT and other OpenAI text generation tools. The reviewed articles
are divided into eleven distinct study topics, organized according to
their contributions and relevance to diverse fields. Furthermore, the
article meticulously compiles and arranges robust statements,
associating them with their relevant domains and citations to provide a
structured and in-depth exploration. In conclusion, the essay highlights
the advantages and disadvantages of ChatGPT and points out areas for
future research to enhance our knowledge and technological prowess.
Overall, this article serves as a valuable resource for those keen on
deepening their understanding of ChatGPT, whether they are in academia,
industry, or technology-driven research initiatives.

\subsection{6.1. Teaching and learning recommendations}

Provide clear guidelines and expectations for university students using
chatbots. Give pupils the knowledge and tools they need to utilize
chatbots ethically and responsibly, emphasizing correct attribution and
moral values {[}143{]}. Educate learners about the advantages of
iteratively interacting with generative AI to enhance their ability to
think critically and reflect in an organized manner {[}144{]}. Utilize
chatbots to create instructional materials such as seminar schedules,
lecture topics, module summaries, announcements, exercises, quizzes, and
other activities {[}145,146{]}. Assist students\textquotesingle{}
learning by offering regular, informal exams and using generative
artificial intelligence to help them apply acquired knowledge to new
situations, question their understanding, and teach them how to evaluate
information critically {[}147{]}. Encourage students to use ChatGPT
critically and analytically, and foster engagement by getting to know
them and showing genuine interest in their ideas {[}147{]}. Clarify
phenomena such as the Eliza effect and explain AI concepts to demystify
the technology. For example, when asked about the connection between
eating worms and making wise life partner selections, ChatGPT (GPT-4
version) provided a clever response relating worm consumption to
open-mindedness and discernment in relationships. This illustrates the
importance, as highlighted by Silva (2023) {[}148{]}, of preparing
students to recognize and challenge nonsense that may sound
authoritative.

\subsection{6.2. Student recommendations}

Recognize the consequences of unethical behavior, understand academic
integrity regulations, use chatbots responsibly, and maintain personal
accountability {[}149{]}. Become proficient with AI tools and develop
digital literacy to improve your career prospects {[}150,151{]}. Rather
than plagiarizing, utilize chatbots for tasks like title creation,
summarizing, and editing, and you can even request that ChatGPT emulate
your favorite author {[}152{]}. Verify the accuracy of your sources and
stay alert to misinformation {[}153,154{]}. Familiarize yourself with a
wide range of reading materials to enhance your analytical and creative
thinking {[}155{]}. Utilize AI-based language tools for coding and
solving real-world problems {[}156{]}. Consider your learning goals and
use AI tools as study partners {[}157{]}. Employ chatbots to condense
lengthy texts, as demonstrated by GPT-4\textquotesingle s success with
works by Goethe and Steinbeck. Lastly, verify all statements made by
chatbots for accuracy and ensure proper referencing, as chatbots may
sometimes mislead.

\subsection{6.3. Higher education recommendations}

Educational institutions are encouraging comprehensive discussions among
various stakeholders, such as students, teachers, faculty from different
disciplines, IT specialists (including those in information systems,
computer science, and data science), career center staff, industry
representatives, legal experts, external authorities (including those
from other higher education institutions), and government officials
{[}158{]}. Furthermore, the insights gained from these dialogues must be
integrated into official policies, guidelines, manuals, and tutorials.
The critical need for digital literacy education, including not only
chatbots but also AI-enhanced tools like Grammarly, must be addressed
through comprehensive digital literacy programs {[}159{]}. Institutions
must also avoid overburdening faculty to ensure that they can engage and
motivate students {[}160{]}.

Training and discussions on AI technologies, such as ChatGPT, are
essential for faculty, while students must be educated about the
importance of maintaining academic integrity when using chatbots
{[}161{]}. There must be a concerted effort to encourage, support, and
disseminate research on the impact of AI technologies on education.
Furthermore, institutions need to update their integrity policies and
honor codes in response to the use of AI tools, creating clear and
comprehensible guidelines for using language models in education and
establishing penalties for cheating {[}162,163{]}.

\subsection{6.4. Future work}

A focus is placed on the effects of text generation, machine learning,
and artificial intelligence (AI) on education in the study. The purpose
of this report is to encourage academic integrity professionals and
educators to embrace rather than reject technology as it is constantly
changing. These technologies will remain based on the same underlying
principles. To encourage a responsible approach to the use of artificial
intelligence in education, the article suggests seven interrelated
topics for investigation. In the first instance, quality control
organizations and academic institutions should develop flexible
regulations to address AI\textquotesingle s role in the classroom while
taking into account its potential future uses. The development of
academic integrity policies requires cooperation. In addition, it is
important to educate students about the proper use of artificial
intelligence, ethical considerations, as well as its limitations, such
as the potential for inaccurate results. Therefore, educational
personnel require assistance in order to comprehend the implications of
artificial intelligence and the necessary adjustments to educational
procedures. It may be necessary to implement a train-the-trainer
strategy due to the lack of experience. Furthermore, different
professions call for different approaches to AI in education, taking
into consideration accreditation criteria, employer needs, and
assessment criteria. Moreover, it is necessary to examine current
evaluation techniques, possibly using artificial intelligence to
facilitate the process or finding alternative strategies to ensure
student participation. The detection mechanisms, including human
detection, must be examined in order to prevent violations of academic
honesty, including possible errors generated by artificial intelligence.
In addition, unions and other international organizations should reflect
the interests of students in the decision-making process.

AI\textquotesingle s contributions explicitly deny authorship of the
document. Since AI is unable to take ownership of the integrity and
content of its work, many organizations agree that AI cannot be
acknowledged as an author. The report emphasizes the limitations of
artificial intelligence, such as its inability to offer current
knowledge, as well as concerns about the spread of pseudoscientific
research that can result from the use of artificial intelligence. It is
important to remember that the Future of Life
Institute\textquotesingle s recent call for an end to AI development
serves as a reminder of the disrupting and hazardous potential of AI,
including the spread of false information and damage to the foundations
of society. As a conclusion, the study reiterates the importance of
moral standards in scholarly publication, pointing out that although AI
may facilitate contributions, it cannot produce reliable ones. It is
essential for the educational community to comprehend, accept, and
properly integrate new technologies despite the challenges and ethical
questions presented by AI\textquotesingle s extraordinary progress. In
this way, teachers may use artificial intelligence to enhance
instruction while upholding academic integrity and preparing their
students for a world where AI will play an increasingly important role
in the future.

\textbf{Funding}

The funding sources had no involvement in the study design, collection,
analysis, or interpretation of data, writing of the manuscript, or in
the decision to submit the manuscript for publication.

\textbf{Competing Interests}

We declare no conflict of interest

\textbf{Data Availability Statement}

This is a review paper, and data availability is not applicable

\textbf{References}

1- Kuleto, V., Ilić, M., Dumangiu, M., Ranković, M., Martins, O. M.,
Păun, D., \& Mihoreanu, L. (2021). Exploring opportunities and
challenges of artificial intelligence and machine learning in higher
education institutions.~Sustainability,~13(18), 10424.

2- Ahmad, S. F., Rahmat, M. K., Mubarik, M. S., Alam, M. M., \& Hyder,
S. I. (2021). Artificial intelligence and its role in
education.~Sustainability,~13(22), 12902.

3- Park, Y. E. (2020). Uncovering trend-based research insights on
teaching and learning in big data. Journal of Big Data, 7(1), 93.

4- Zhai, X., Chu, X., Chai, C. S., Jong, M. S. Y., Istenic, A., Spector,
M., ... \& Li, Y. (2021). A Review of Artificial Intelligence (AI) in
Education from 2010 to 2020. Complexity, 2021, 1-18.

5- Fuchs, K. (2023, May). Exploring the opportunities and challenges of
NLP models in higher education: is Chat GPT a blessing or a curse?. In
Frontiers in Education (Vol. 8, p. 1166682). Frontiers.

6- Jiang, Q., Zhang, Y., \& Pian, W. (2022). Chatbot as an emergency
exist: Mediated empathy for resilience via human-AI interaction during
the COVID-19 pandemic.~Information processing \& management,~59(6),
103074.

7- Sandu, N., \& Gide, E. (2019, September). Adoption of AI-Chatbots to
enhance student learning experience in higher education in India. In
2019 18th International Conference on Information Technology Based
Higher Education and Training (ITHET) (pp. 1-5). IEEE.

8- Seo, K., Tang, J., Roll, I., Fels, S., \& Yoon, D. (2021). The impact
of artificial intelligence on learner--instructor interaction in online
learning.~International journal of educational technology in higher
education,~18(1), 1-23.

9- Sarker, I. H. (2021). Deep learning: a comprehensive overview on
techniques, taxonomy, applications and research directions.~SN Computer
Science,~2(6), 420.

10- Clarke R, Lancaster T (2006) Eliminating the successor to
plagiarism? Identifying the usage of contract cheating sites.
Proceedings of 2nd International Plagiarism Conference. Newcastle,
United Kingdom

11- Kuleto, V., Ilić, M., Dumangiu, M., Ranković, M., Martins, O. M.,
Păun, D., \& Mihoreanu, L. (2021). Exploring opportunities and
challenges of artificial intelligence and machine learning in higher
education institutions. Sustainability, 13(18), 10424.

12- Kung, T. H., Cheatham, M., Medenilla, A., Sillos, C., De Leon, L.,
Elepaño, C., ... \& Tseng, V. (2023). Performance of ChatGPT on USMLE:
Potential for AI-assisted medical education using large language
models.~PLoS digital health,~2(2), e0000198.

13- Keiper, M. C. (2023). ChatGPT in practice: Increasing event planning
efficiency through artificial intelligence. Journal of Hospitality,
Leisure, Sport \& Tourism Education, 33, 100454.

14- Gozalo-Brizuela, R., \& Garrido-Merchan, E. C. (2023). ChatGPT is
not all you need. A State of the Art Review of large Generative AI
models. arXiv preprint arXiv:2301.04655.

15- Chrisinger, B. W. (2023). It\textquotesingle s not just our
students---ChatGPT is coming for faculty writing.

16- Roe J, Perkins M (2022) What are Automated Paraphrasing Tools and
how do we address them? A review of a growing threat to academic
integrity. Int J Educ Integr 18:15.
\url{https://doi.org/10.1007/s40979-022-00109-w}

17- Jalil, S., Rafi, S., LaToza, T. D., Moran, K., \& Lam, W. (2023,
April). Chatgpt and software testing education: Promises \& perils.
In~2023 IEEE International Conference on Software Testing, Verification
and Validation Workshops (ICSTW)~(pp. 4130-4137). IEEE.

18- AYDIN, Ö., \& KARAARSLAN, E. (2023). Is chatgpt leading generative
ai? what is beyond expectations.~What is Beyond Expectations.

19- West, D. M., \& Allen, J. R. (2018). How artificial intelligence is
transforming the world.~Report. April,~24, 2018.

20- Benbya, H., Davenport, T. H., \& Pachidi, S. (2020). Artificial
intelligence in organizations: Current state and future
opportunities.~MIS Quarterly Executive,~19(4).

21- Fitria, T. N. (2021). The use technology based on artificial
intelligence in English teaching and learning.~ELT Echo: The Journal of
English Language Teaching in Foreign Language Context,~6(2), 213-223.

22- Adamopoulou, E., \& Moussiades, L. (2020). Chatbots: History,
technology, and applications.~Machine Learning with Applications,~2,
100006.

23- Skrebeca, J., Kalniete, P., Goldbergs, J., Pitkevica, L.,
Tihomirova, D., \& Romanovs, A. (2021, October). Modern development
trends of chatbots using artificial intelligence (ai). In~2021 62nd
International Scientific Conference on Information Technology and
Management Science of Riga Technical University (ITMS)~(pp. 1-6). IEEE.

24- Suta, P., Lan, X., Wu, B., Mongkolnam, P., \& Chan, J. H. (2020). An
overview of machine learning in chatbots.~International Journal of
Mechanical Engineering and Robotics Research,~9(4), 502-510.

25- Mariacher, N., Schlögl, S., \& Monz, A. (2021). Investigating
perceptions of social intelligence in simulated human-chatbot
interactions.~Progresses in Artificial Intelligence and Neural Systems,
513-529.

26- Cassell, J. (2001). Embodied conversational agents: representation
and intelligence in user interfaces.~AI magazine,~22(4), 67-67.

27- Cooper, B., Brna, P., \& Martins, A. (1999, October). Effective
affective in intelligent systems--building on evidence of empathy in
teaching and learning. In~International Workshop on Affective
Interactions~(pp. 21-34). Berlin, Heidelberg: Springer Berlin
Heidelberg.

28- Lund, B. D., Wang, T., Mannuru, N. R., Nie, B., Shimray, S., \&
Wang, Z. (2023). ChatGPT and a new academic reality: Artificial
Intelligence‐written research papers and the ethics of the large
language models in scholarly publishing.~Journal of the Association for
Information Science and Technology,~74(5), 570-581.

29- Parikh, P. M., Shah, D. M., Parikh, U. G., Venniyoor, A., Babu, G.,
Garg, A., \& Malhotra, H. (2023). ChatGPT---Preliminary Overview with
Implications for Medicine and Oncology.~Indian Journal of Medical and
Paediatric Oncology.

30- Peng, Z., Wang, X., Han, Q., Zhu, J., Ma, X., \& Qu, H. (2023).
Storyfier: Exploring Vocabulary Learning Support with Text Generation
Models.~arXiv preprint arXiv:2308.03864.

31- Panda, S., \& Kaur, N. (2023). Exploring the viability of ChatGPT as
an alternative to traditional chatbot systems in library and information
centers.~Library Hi Tech News,~40(3), 22-25.

32- Fitria, T. N. (2023, March). Artificial intelligence (AI) technology
in OpenAI ChatGPT application: A review of ChatGPT in writing English
essay. In~ELT Forum: Journal of English Language Teaching~(Vol. 12, No.
1, pp. 44-58).

33- Rahman, A. M., Al Mamun, A., \& Islam, A. (2017, December).
Programming challenges of chatbot: Current and future prospective.
In~2017 IEEE region 10 humanitarian technology conference (R10-HTC)~(pp.
75-78). IEEE.

34- Zhou, C., Li, Q., Li, C., Yu, J., Liu, Y., Wang, G., ... \& Sun, L.
(2023). A comprehensive survey on pretrained foundation models: A
history from bert to chatgpt. arXiv preprint arXiv:2302.09419.

35- Haluza, D., \& Jungwirth, D. (2023). Artificial Intelligence and Ten
Societal Megatrends: An Exploratory Study Using GPT-3.~Systems,~11(3),
120.

36- Sohail, S. S., Farhat, F., Himeur, Y., Nadeem, M., Madsen, D. Ø.,
Singh, Y., ... \& Mansoor, W. (2023). Decoding ChatGPT: A Taxonomy of
Existing Research, Current Challenges, and Possible Future
Directions.~Journal of King Saud University-Computer and Information
Sciences, 101675.

37- Javaid, M., Haleem, A., \& Singh, R. P. (2023). ChatGPT for
healthcare services: An emerging stage for an innovative perspective.
BenchCouncil Transactions on Benchmarks, Standards and Evaluations,
3(1), 100105.

38- Nastasi, A.J.; Courtright, K.R.; Halpern, S.D.; Weissman, G.E. Does
Chatgpt Provide Appropriate and Equitable Medical Advice? A
Vignette-Based, Clinical Evaluation Across Care Contexts. medRxiv 2023.

39- Grünebaum, A., Chervenak, J., Pollet, S. L., Katz, A., \& Chervenak,
F. A. (2023). The exciting potential for ChatGPT in obstetrics and
gynecology. American Journal of Obstetrics and Gynecology, 228(6),
696-705.

40- Roumeliotis, K. I., \& Tselikas, N. D. (2023). ChatGPT and Open-AI
Models: A Preliminary Review.~Future Internet,~15(6), 192.

41- Birenbaum, M. (2023). The Chatbots' Challenge to Education:
Disruption or Destruction?.~Education Sciences,~13(7), 711.

42- Roumeliotis, K. I., \& Tselikas, N. D. (2023). ChatGPT and Open-AI
Models: A Preliminary Review.~Future Internet,~15(6), 192.

43-Raković, M., Sha, L., Nagtzaam, G., Young, N., Stratmann, P.,
Gašević, D., \& Chen, G. (2022, July). Towards the automated evaluation
of legal casenote essays. In~International Conference on Artificial
Intelligence in Education~(pp. 167-179). Cham: Springer International
Publishing.

44-Zhong, T., Wei, Y., Yang, L., Wu, Z., Liu, Z., Wei, X., ... \& Zhang,
T. (2023). Chatabl: Abductive learning via natural language interaction
with chatgpt.~arXiv preprint arXiv:2304.11107.

45- Ma, W., Liu, S., Wang, W., Hu, Q., Liu, Y., Zhang, C., ... \& Liu,
Y. (2023). The Scope of ChatGPT in Software Engineering: A Thorough
Investigation.~arXiv preprint arXiv:2305.12138.

46- Dergaa, I., Chamari, K., Zmijewski, P., \& Saad, H. B. (2023). From
human writing to artificial intelligence generated text: examining the
prospects and potential threats of ChatGPT in academic writing. Biology
of Sport, 40(2), 615-622.

47- Zhong, Q., Ding, L., Liu, J., Du, B., \& Tao, D. (2023). Can chatgpt
understand too? a comparative study on chatgpt and fine-tuned
bert.~arXiv preprint arXiv:2302.10198.

48- Mogavi, R. H., Deng, C., Kim, J. J., Zhou, P., Kwon, Y. D.,
Metwally, A. H. S., ... \& Hui, P. (2023). Exploring user perspectives
on chatgpt: Applications, perceptions, and implications for
ai-integrated education.~arXiv preprint arXiv:2305.13114.

49-Guo, B., Zhang, X., Wang, Z., Jiang, M., Nie, J., Ding, Y., ... \&
Wu, Y. (2023). How close is chatgpt to human experts? comparison corpus,
evaluation, and detection. arXiv preprint arXiv:2301.07597.

50- Rahman, M. M., \& Watanobe, Y. (2023). ChatGPT for education and
research: Opportunities, threats, and strategies.~Applied
Sciences,~13(9), 5783.

51- Liu, Y., Han, T., Ma, S., Zhang, J., Yang, Y., Tian, J., ... \& Ge,
B. (2023). Summary of chatgpt/gpt-4 research and perspective towards the
future of large language models.~arXiv preprint arXiv:2304.01852.

52- Hassani, H., \& Silva, E. S. (2023). The role of ChatGPT in data
science: how ai-assisted conversational interfaces are revolutionizing
the field.~Big data and cognitive computing,~7(2), 62.

53- Lin, C. C., Huang, A. Y., \& Yang, S. J. (2023). A review of
ai-driven conversational chatbots implementation methodologies and
challenges (1999--2022).~Sustainability,~15(5), 4012.

54- Zielinski, C., Winker, M., Aggarwal, R., Ferris, L., Heinemann, M.,
Lapeña, J. F., ... \& Citrome, L. (2023). Chatbots, ChatGPT, and
Scholarly Manuscripts-WAME Recommendations on ChatGPT and Chatbots in
Relation to Scholarly Publications. Afro-Egyptian Journal of Infectious
and Endemic Diseases, 13(1), 75-79.

55- Rudolph, J., Tan, S., \& Tan, S. (2023). ChatGPT: Bullshit spewer or
the end of traditional assessments in higher education?. Journal of
Applied Learning and Teaching, 6(1).

56- Susnjak, T.: Applying bert and chatgpt for sentiment analysis of
lyme disease in scientific literature. arXiv preprint arXiv:2302.06474
(2023)

57- Tlili, A., Shehata, B., Adarkwah, M. A., Bozkurt, A., Hickey, D. T.,
Huang, R., \& Agyemang, B. (2023). What if the devil is my guardian
angel: ChatGPT as a case study of using chatbots in education.~Smart
Learning Environments,~10(1), 15.

58- Kovačević, D. Use of chatgpt in ESP teaching process. In Proceedings
of the 2023 22nd International Symposium INFOTEH-JAHORINA (INFOTEH),
East Sarajevo, Bosnia and Herzegovina, 15--17 March 2023. {[}Google
Scholar{]}

59- Shoufan, A. Exploring Students' Perceptions of CHATGPT: Thematic
Analysis and Follow-Up Survey; IEEE Access: New York, NY, USA, 2023; pp.
38805--38818.

60- Castellanos-Gomez, A. Good Practices for Scientific Article Writing
with ChatGPT and Other Artificial Intelligence Language Models.
Nanomanufacturing 2023, 3, 135--138.

61- Murad, D.F.; Heryadi, Y.; Wijanarko, B.D.; Isa, S.M.; Budiharto, W.
Recommendation system for smart LMS using machine learning: A literature
review. In Proceedings of the 2018 International Conference on
Computing, Engineering, and Design (ICCED), Bangkok, Thailand, 6--8
September 2018; pp. 113--118.

62- Sciarrone, F. Machine learning and learning analytics: Integrating
data with learning. In Proceedings of the 2018 17th International
Conference on Information Technology Based Higher Education and Training
(ITHET), Olhao, Portugal, 26--28 April 2018; pp. 1--5.

63- Romero, C.; Ventura, S. Educational Data Mining: A Review of the
State of the Art. IEEE Trans. Syst. Man, Cybern. Part C (Appl. Rev.)
2010, 40, 601--618.

64- Maddigan, P., \& Susnjak, T. (2023). Chat2vis: Generating data
visualisations via natural language using chatgpt, codex and gpt-3 large
language models.~IEEE Access.

65-Liu, X., He, P., Chen, W., \& Gao, J. (2019). Improving multi-task
deep neural networks via knowledge distillation for natural language
understanding.~arXiv preprint arXiv:1904.09482.

66- Ahmad, A., Waseem, M., Liang, P., Fahmideh, M., Aktar, M. S., \&
Mikkonen, T. (2023, June). Towards human-bot collaborative software
architecting with chatgpt. In Proceedings of the 27th International
Conference on Evaluation and Assessment in Software Engineering (pp.
279-285).

67- Norouzi, S., \& Ebrahimi, M. (2019). A Survey on Proposed Methods to
Address Adam Optimizer Deficiencies.

68- Wang, R., \& Klabjan, D. (2022). Divergence results and convergence
of a variance reduced version of adam.~arXiv preprint arXiv:2210.05607.

69- Rahaman, M. S., Ahsan, M. M., Anjum, N., Rahman, M. M., \& Rahman,
M. N. (2023). The AI race is on! Google\textquotesingle s Bard and
OpenAI\textquotesingle s ChatGPT head to head: an opinion
article.~Mizanur and Rahman, Md Nafizur, The AI Race is on.

70- Pavlik, J. V. (2023). Collaborating with ChatGPT: Considering the
implications of generative artificial intelligence for journalism and
media education.~Journalism \& Mass Communication Educator,~78(1),
84-93.

71- Yu, H. (2023). Reflection on whether Chat GPT should be banned by
academia from the perspective of education and teaching.~Frontiers in
Psychology,~14, 1181712.

72- Veselovsky, V., Ribeiro, M. H., \& West, R. (2023). Artificial
Artificial Artificial Intelligence: Crowd Workers Widely Use Large
Language Models for Text Production Tasks.~arXiv preprint
arXiv:2306.07899.

73- Dao, X. Q. (2023). Performance comparison of large language models
on vnhsge english dataset: Openai chatgpt, microsoft bing chat, and
google bard. arXiv preprint arXiv:2307.02288.

74- Ribas, J. (2023a, February 22). Building the new Bing. \RL{}

75- Guo, Y., \& Yu, H. Exploring Project-Based Learning in English
Teaching for the Development of Core Literacy in Primary and Secondary
Education: Effective Strategies to Enhance Students\textquotesingle{}
Comprehensive Skills.

76- Iqbal, M., Khalid, M. N., Manzoor, A., Abid, M. M., \& Shaikh, N. A.
(2022). Search Engine Optimization (SEO): A Study of important key
factors in achieving a better Search Engine Result Page (SERP)
Position.~Sukkur IBA Journal of Computing and Mathematical
Sciences,~6(1), 1-15.

77- Vasconcelos, M. A. R., \& Santos, R. P. D. (2023). Enhancing STEM
learning with ChatGPT and Bing Chat as objects to think with: A case
study. arXiv preprint arXiv:2305.02202. 78-79-

78- Dao, X. Q. (2023). Which Large Language Model should You Use in
Vietnamese Education: ChatGPT, Bing Chat, or Bard?. Bing Chat, or Bard

79-. Xuan-Quy, D., Ngoc-Bich, L., Xuan-Dung, P., Bac-Bien, N., \&
The-Duy, V. (2023). Evaluation of ChatGPT and Microsoft Bing AI Chat
Performances on Physics Exams of Vietnamese National High School
Graduation Examination. arXiv preprint arXiv:2306.04538.

80- Dhanvijay, A. K. D., Pinjar, M. J., Dhokane, N., Sorte, S. R.,
Kumari, A., Mondal, H., \& Dhanvijay, A. K. (2023). Performance of Large
Language Models (ChatGPT, Bing Search, and Google Bard) in Solving Case
Vignettes in Physiology. Cureus, 15(8).

81- Koga, S., Martin, N. B., \& Dickson, D. W. (2023). Evaluating the
performance of large language models: ChatGPT and Google Bard in
generating differential diagnoses in clinicopathological conferences of
neurodegenerative disorders. Brain Pathology, e13207.

82- Santos, R. P. D. (2023). Enhancing Physics Learning with ChatGPT,
Bing Chat, and Bard as Agents-to-Think-With: A Comparative Case Study.
arXiv preprint arXiv:2306.00724.

83- AYDIN, Ö. (2023). Google Bard generated literature review:
metaverse. Journal of AI, 7(1), 1-14.

84- Google Bard: Key Features, Best Uses, Architecture, Important
Limitations - factigator.com. (2023, February 17). factigator.com -
Factual Information You Can Rely On.
https://factigator.com/google-bard-key-features-best-uses-architecture-important-limitations/

85- Chen, B. X., Grant, N., \& Weise, K. (2023, March 15). How Siri,
Alexa and Google Assistant lost the A.I. race. The New York Times,
https://www.nytimes.com/2023/03/15/technology/
siri-alexa-google-assistant-artificial-intelligence.html?a
ction=click\&pgtpe=Article\&state=default\&module=sty
ln-artificial-intelligence\&variant=show\&region=BELOW\_
MAIN\_CONTENT\&block=storyline\_flex\_guide\_recirc

86- Ye, J. (2023, June 27). China's Baidu says its new AI beat ChatGPT
on some metrics. Reuters.
https://www.reuters.com/technology/chinas-baidu-says-its-new-ai-beat-chatgpt-some-metrics-2023-06-27/

87- Chinese tech giant Baidu just released its answer to ChatGPT. (2023,
March 16). MIT Technology Review.
https://www.technologyreview.com/2023/03/16/1069919/baidu-ernie-bot-chatgpt-launch/

88- Baron, N. S. (2023). Who Wrote This?: How AI and the Lure of
Efficiency Threaten Human Writing. Stanford University Press.

89- Sun, Y., Wang, S., Feng, S., Ding, S., Pang, C., Shang, J., ... \&
Wang, H. (2021). Ernie 3.0: Large-scale knowledge enhanced pre-training
for language understanding and generation. arXiv preprint
arXiv:2107.02137.

90- Baptista, E., \& Ye, J. (2023). China's answer to ChatGPT? Baidu
shares tumble as Ernie Bot disappoints. Reuters, https://
\href{http://www.reuters.com/technology/chinese-search-giant-baiduintroduces-ernie-bot-2023-03-16/}{www.reuters.com/technology/chinese-search-giant-baiduintroduces-ernie-bot-2023-03-16/}

91- Yang, Z. (2023b, March 16). Chinese tech giant Baidu justbreleased
its answer to ChatGPT. MIT Technology
Review,nhttps://www.technologyreview.com/2023/03/16/1069919/baidu-ernie-bot-chatgpt-launch/

92- Lund, B. D., \& Wang, T. (2023). Chatting about ChatGPT: how may AI
and GPT impact academia and libraries?. Library Hi Tech News, 40(3),
26-29.

93- Qammar, A., Wang, H., Ding, J., Naouri, A., Daneshmand, M., \& Ning,
H. (2023). Chatbots to ChatGPT in a Cybersecurity Space: Evolution,
Vulnerabilities, Attacks, Challenges, and Future Recommendations. arXiv
preprint arXiv:2306.09255.

94- Singh, V. (2023). Unleashing AI: a partner, a deferent or a
confining force?. Amazon.

95- Firat, M. (2023). What ChatGPT means for universities: Perceptions
of scholars and students. Journal of Applied Learning and Teaching,
6(1).

96- Li, M., Erickson, I. M., Cross, E. V., \& Lee, J. D. (2023). It's
Not Only What You Say, But Also How You Say It: Machine Learning
Approach to Estimate Trust from Conversation.~Human Factors,
00187208231166624.

97- Zhai, X. (2023). Chatgpt for next generation science learning.~XRDS:
Crossroads, The ACM Magazine for Students,~29(3), 42-46.

98- Biswas, S. (2023). Role of Chat GPT in Education.~Available at SSRN
4369981.

99- Liu, J., \& Liu, S. (2023). The application of ChatGPT in medical
education.

100- Halaweh, M. (2023). ChatGPT in education: Strategies for
responsible implementation.

101- Susnjak, T. (2022). ChatGPT: The end of online exam integrity?.
arXiv preprint arXiv:2212.09292.

102- McMurtrie, B. (2022, December 13). AI and the future of
undergraduate writing. The Chronicle of Higher Education.
https://www.chronicle.com/article/ai-and-the-future-ofundergraduate-writing

103- Sharples, M. (2022). Automated essay writing: An AIED opinion.
International journal of artificial intelligence in education, 32(4),
1119-1126.

104- Qadir, J. (2023, May). Engineering education in the era of ChatGPT:
Promise and pitfalls of generative AI for education. In 2023 IEEE Global
Engineering Education Conference (EDUCON) (pp. 1-9). IEEE.

105- Bozkurt, A., Xiao, J., Lambert, S., Pazurek, A., Crompton, H.,
Koseoglu, S., ... \& Jandrić, P. (2023). Speculative futures on ChatGPT
and generative artificial intelligence (AI): A collective reflection
from the educational landscape. Asian Journal of Distance Education,
18(1).

106- Mhlanga, D. (2023). Open AI in education, the responsible and
ethical use of ChatGPT towards lifelong learning. Education, the
Responsible and Ethical Use of ChatGPT Towards Lifelong Learning
(February 11, 2023).

107- Adiguzel, T., Kaya, M. H., \& Cansu, F. K. (2023). Revolutionizing
education with AI: Exploring the transformative potential of ChatGPT.
Contemporary Educational Technology, 15(3), ep429.

108- Elbanna, S., \& Armstrong, L. (2023). Exploring the integration of
ChatGPT in education: adapting for the future. Management \&
Sustainability: An Arab Review..

109- Sullivan, M., Kelly, A., \& McLaughlan, P. (2023). ChatGPT in
higher education: Considerations for academic integrity and student
learning.

110- Haensch, A. C., Ball, S., Herklotz, M., \& Kreuter, F. (2023).
Seeing ChatGPT Through Students\textquotesingle{} Eyes: An Analysis of
TikTok Data.~arXiv preprint arXiv:2303.05349.

111- Kasneci, E., Seßler, K., Küchemann, S., Bannert, M., Dementieva,
D., Fischer, F., ... \& Kasneci, G. (2023). ChatGPT for good? On
opportunities and challenges of large language models for
education.~Learning and Individual Differences,~103, 102274.

112- Henriksen, D., Woo, L. J., \& Mishra, P. (2023). Creative Uses of
ChatGPT for Education: a Conversation with Ethan Mollick.~TechTrends,
1-6.

113- Mollick, E. R., \& Mollick, L. (2022). New modes of learning
enabled by ai chatbots: Three methods and assignments.~Available at
SSRN.

114- Gimpel, H., Hall, K., Decker, S., Eymann, T., Lämmermann, L.,
Mädche, A., ... \& Vandrik, S. (2023).~Unlocking the power of generative
AI models and systems such as GPT-4 and ChatGPT for higher education: A
guide for students and lecturers~(No. 02-2023). Hohenheim Discussion
Papers in Business, Economics and Social Sciences.

115- Nisar, S., \& Aslam, M. S. (2023). Is ChatGPT a Good Tool for T\&CM
Students in Studying Pharmacology?.~Available at SSRN 4324310.

116- Lo, C. K. (2023). What is the impact of ChatGPT on education? A
rapid review of the literature.~Education Sciences,~13(4), 410.

117- Ray, P. P. (2023). ChatGPT: A comprehensive review on background,
applications, key challenges, bias, ethics, limitations and future
scope. Internet of Things and Cyber-Physical Systems.

118- Sallam, M. (2023, March). ChatGPT utility in healthcare education,
research, and practice: systematic review on the promising perspectives
and valid concerns. In~Healthcare~(Vol. 11, No. 6, p. 887). MDPI.

119-Hargreaves, S. (2023). `Words Are Flowing Out Like Endless Rain Into
a Paper Cup': ChatGPT \& Law School Assessments. The Chinese University
of Hong Kong Faculty of Law Research Paper, (2023-03).

120-Dai, W., Lin, J., Jin, F., Li, T., Tsai, Y. S., Gasevic, D., \&
Chen, G. (2023). Can large language models provide feedback to students?
A case study on ChatGPT.

121- Crawford, J., Cowling, M., Ashton-Hay, S., Kelder, J. A.,
Middleton, R., \& Wilson, G. S. (2023). Artificial Intelligence and
Authorship Editor Policy: ChatGPT, Bard Bing AI, and beyond.~Journal of
University Teaching \& Learning Practice,~20(5), 1.

122- Crawford, J., Cowling, M., \& Allen, K. A. (2023). Leadership is
needed for ethical ChatGPT: Character, assessment, and learning using
artificial intelligence (AI).~Journal of University Teaching \& Learning
Practice,~20(3), 02.

123- Baidoo-Anu, D., \& Owusu Ansah, L. (2023). Education in the era of
generative artificial intelligence (AI): Understanding the potential
benefits of ChatGPT in promoting teaching and learning.~Available at
SSRN 4337484.

124- Grassini, S. (2023). Shaping the future of education: exploring the
potential and consequences of AI and ChatGPT in educational settings.
Education Sciences, 13(7), 692.

125- Felten, E., Raj, M., \& Seamans, R. (2023). How will Language
Modelers like ChatGPT Affect Occupations and Industries?.~arXiv preprint
arXiv:2303.01157.

126- Pfeffer, O. P., Sailer, M., Schmidt, A., Seidel, T., Stadler, M.,
Weller, J., ... \& Kasneci, G. (2023). ChatGPT for Good? On
Opportunities and Challenges of Large Language Models for Education.
2023-01-15{]}.

127- Fui-Hoon Nah, F., Zheng, R., Cai, J., Siau, K., \& Chen, L. (2023).
Generative AI and ChatGPT: Applications, challenges, and AI-human
collaboration. Journal of Information Technology Case and Application
Research, 1-28.

128- Peters, M. A., Jackson, L., Papastephanou, M., Jandrić, P.,
Lazaroiu, G., Evers, C. W., ... \& Fuller, S. (2023). AI and the future
of humanity: ChatGPT-4, philosophy and education--Critical responses.
Educational Philosophy and Theory, 1-35.

129- Griffith, E., \& Metz, C. (2023, March 14). `Let 1,000 flowers
bloom': A.I. funding frenzy escalates. The New York Times,
\url{https://www}.
nytimes.com/2023/03/14/technology/ai-funding-boom.html?action=click\&module=RelatedLinks\&pgtype=Article

130- Lee, K.-F. (2018). AI superpowers. China, Silicon Valley and the
new world order. Houghton Mifflin Harcourt.

131- Yang, L., \& Wang, J. (2023). Factors influencing initial public
acceptance of integrating the ChatGPT-type model with government
services. Kybernetes.

132- Hew, K. F., Huang, W., Du, J., \& Jia, C. (2023). Using chatbots to
support student goal setting and social presence in fully online
activities: learner engagement and perceptions. Journal of Computing in
Higher Education, 35(1), 40-68.

133- Schäfer, M. S. (2023). The Notorious GPT: science communication in
the age of artificial intelligence.~Journal of Science
Communication,~22(2), Y02.

134- Rudolph, J., Tan, S., \& Tan, S. (2023). War of the chatbots: Bard,
Bing Chat, ChatGPT, Ernie and beyond. The new AI gold rush and its
impact on higher education.~Journal of Applied Learning and
Teaching,~6(1).

135- Santos, R. P. D. (2023). Enhancing Physics Learning with ChatGPT,
Bing Chat, and Bard as Agents-to-Think-With: A Comparative Case
Study.~arXiv preprint arXiv:2306.00724.

136- Dao, X. Q., \& Le, N. B. (2023). Chatgpt is good but bing chat is
better for vietnamese students.~arXiv preprint arXiv:2307.08272.

137- Lin, J. (2023). ChatGPT and Moodle Walk into a Bar: A Demonstration
of AI's Mind-blowing Impact on E-Learning. Available at SSRN.

138- Balaji, S., Rajaram, V., \& Sethu, S. (2023, April). Gamification
of Learning using 3D Spaces in the Metaverse. In 2023 International
Conference on Networking and Communications (ICNWC) (pp. 1-5). IEEE.

139- Thu, C. H., Bang, H. C., \& Cao, L. (2023). Integrating ChatGPT
into Online Education System in Vietnam: Opportunities and Challenges.

140- Zhai, X. (2022). ChatGPT user experience: Implications for
education. Available at SSRN 4312418.

141- Rasul, T., Nair, S., Kalendra, D., Robin, M., de Oliveira Santini,
F., Ladeira, W. J., ... \& Heathcote, L. (2023). The role of ChatGPT in
higher education: Benefits, challenges, and future research directions.
Journal of Applied Learning and Teaching, 6(1).

142- Haleem, A., Javaid, M., \& Singh, R. P. (2022). An era of ChatGPT
as a significant futuristic support tool: A study on features,
abilities, and challenges. BenchCouncil transactions on benchmarks,
standards and evaluations, 2(4), 100089.

143- Kohnke, L., Moorhouse, B. L., \& Zou, D. (2023). ChatGPT for
language teaching and learning. RELC Journal, 00336882231162868.

144- Opara, E., Mfon-Ette Theresa, A., \& Aduke, T. C. (2023). ChatGPT
for teaching, learning and research: Prospects and challenges. Opara
Emmanuel Chinonso, Adalikwu Mfon-Ette Theresa, Tolorunleke Caroline
Aduke (2023). ChatGPT for Teaching, Learning and Research: Prospects and
Challenges. Glob Acad J Humanit Soc Sci, 5.

145- Nov, O., Singh, N., \& Mann, D. (2023). Putting ChatGPT's Medical
Advice to the (Turing) Test: Survey Study. JMIR Medical Education, 9(1),
e46939.

146- Almusaed, A., Almssad, A., Yitmen, I., \& Homod, R. Z. (2023).
Enhancing Student Engagement: Harnessing ``AIED''\,'s Power in Hybrid
Education---A Review Analysis. Education Sciences, 13(7), 632.

147- Barter, D. (2023). Degrees of Freedom: Designing Information and
Communication Technologies to Support Enhanced Agency for Blind and
Partially Sighted Individuals Through Cross-Sensory Information
Representation.

148- De Silva, D., Mills, N., El-Ayoubi, M., Manic, M., \& Alahakoon, D.
(2023, April). ChatGPT and Generative AI Guidelines for Addressing
Academic Integrity and Augmenting Pre-Existing Chatbots. In~2023 IEEE
International Conference on Industrial Technology (ICIT)~(pp. 1-6).
IEEE.

149- AlAfnan, M. A., Dishari, S., Jovic, M., \& Lomidze, K. (2023).
Chatgpt as an educational tool: Opportunities, challenges, and
recommendations for communication, business writing, and composition
courses. Journal of Artificial Intelligence and Technology, 3(2), 60-68.

150- Zhang, P. (2023). Taking Advice from ChatGPT. arXiv preprint
arXiv:2305.11888.

151- Castonguay, A., Farthing, P., Davies, S., Vogelsang, L., Kleib, M.,
Risling, T., \& Green, N. (2023). Revolutionizing nursing education
through Ai integration: A reflection on the disruptive impact of
ChatGPT. Nurse Education Today, 129, 105916.

152- Javaid, M., Haleem, A., Singh, R. P., Khan, S., \& Khan, I. H.
(2023). Unlocking the opportunities through ChatGPT Tool towards
ameliorating the education system. BenchCouncil Transactions on
Benchmarks, Standards and Evaluations, 3(2), 100115.

153- Farrokhnia, M., Banihashem, S. K., Noroozi, O., \& Wals, A. (2023).
A SWOT analysis of ChatGPT: Implications for educational practice and
research. Innovations in Education and Teaching International, 1-15.

154- Oviedo-Trespalacios, O., Peden, A. E., Cole-Hunter, T., Costantini,
A., Haghani, M., Rod, J. E., ... \& Reniers, G. (2023). The risks of
using chatgpt to obtain common safety-related information and advice.
Safety Science, 167, 106244.

155- Gill, S. S., Xu, M., Patros, P., Wu, H., Kaur, R., Kaur, K., ... \&
Buyya, R. (2024). Transformative effects of ChatGPT on modern education:
Emerging Era of AI Chatbots. Internet of Things and Cyber-Physical
Systems, 4, 19-23.

156- Berşe, S., Akça, K., Dirgar, E., \& Kaplan Serin, E. (2023). The
role and potential contributions of the artificial intelligence language
model ChatGPT. Annals of Biomedical Engineering, 1-4.

157- Ahmed, I., Kajol, M., Hasan, U., Datta, P. P., Roy, A., \& Reza, M.
R. (2023). ChatGPT vs. Bard: A Comparative Study. UMBC Student
Collection.

158- Newton, P. M., \& Xiromeriti, M. (2023). ChatGPT Performance on MCQ
Exams in Higher Education. A Pragmatic Scoping Review.

159- Jamal, A., Solaiman, M., Alhasan, K., Temsah, M. H., \& Sayed, G.
(2023). Integrating ChatGPT in Medical Education: Adapting Curricula to
Cultivate Competent Physicians for the AI Era. Cureus, 15(8).

160- Bin-Hady, W. R. A., Al-Kadi, A., Hazaea, A., \& Ali, J. K. M.
(2023). Exploring the dimensions of ChatGPT in English language
learning: A global perspective. Library Hi Tech.

161- Schönberger, M. ChatGPT in higher education: the good, the bad, and
the University.

162- Oyelude, A. A. (2023). Much ado about ChatGPT: libraries and
librarians perspectives. Library Hi Tech News, 40(3), 15-17.

163- Tsai, M. L., Ong, C. W., \& Chen, C. L. (2023). Exploring the use
of large language models (LLMs) in chemical engineering education:
Building core course problem models with Chat-GPT. Education for
Chemical Engineers, 44, 71-95.

164- Vázquez-Cano, E., Ramírez-Hurtado, J. M., Sáez-López, J. M., \&
López-Meneses, E. (2023). ChatGPT: The Brightest Student in the
Class.~Thinking Skills and Creativity, 101380.

165- Jahic, I., Ebner, M., \& Schön, S. (2023). Harnessing the power of
artificial intelligence and ChatGPT in education--a first rapid
literature review. Proceedings of EdMedia+ Innovate Learning 2023,
1462-1470.

166- Bitzenbauer, P. (2023). ChatGPT in physics education: A pilot study
on easy-to-implement activities.~Contemporary Educational
Technology,~15(3), ep430.

167- Ryznar, M. (2023). Exams in the Time of ChatGPT.~Washington and Lee
Law Review Online,~80(5), 305.

168- Wardat, Y., Tashtoush, M. A., AlAli, R., \& Jarrah, A. M. (2023).
ChatGPT: A revolutionary tool for teaching and learning mathematics.
Eurasia Journal of Mathematics, Science and Technology Education, 19(7),
em2286.

169- Hwang, A., Oza, N., Callison-Burch, C., \& Head, A. (2023).
Rewriting the Script: Adapting Text Instructions for Voice
Interaction.~arXiv preprint arXiv:2306.09992.

170-Dwivedi, Y. K., Kshetri, N., Hughes, L., Slade, E. L., Jeyaraj, A.,
Kar, A. K., ... \& Wright, R. (2023). ``So what if ChatGPT wrote it?''
Multidisciplinary perspectives on opportunities, challenges and
implications of generative conversational AI for research, practice and
policy. International Journal of Information Management, 71, 102642.

171- Crompton, H., \& Burke, D. (2023). Artificial intelligence in
higher education: the state of the field. International Journal of
Educational Technology in Higher Education, 20(1), 1-22.

172. Huallpa, J. J. (2023). Exploring the ethical considerations of
using Chat GPT in university education.~Periodicals of Engineering and
Natural Sciences,~11(4), 105-115.

173. HEUNG, Y. M. E., \& Chiu, T. K. (2025). How ChatGPT impacts student
engagement from a systematic review and meta-analysis study. Computers
and Education: Artificial Intelligence, 100361.

174. Maaß, L., Grab-Kroll, C., Koerner, J., Öchsner, W., Schön, M.,
Messerer, D. A. C., ... \& Böckers, A. (2025). Artificial Intelligence
and ChatGPT in Medical Education: A Cross-Sectional Questionnaire on
students' Competence. Journal of CME, 14(1), 2437293.

175. Mustofa, H. A., Kola, A. J., \& Owusu-Darko, I. (2025). Integration
of Artificial Intelligence (ChatGPT) into Science Teaching and Learning.
International Journal of Ethnoscience and Technology in Education, 2(1),
108-128.

176. Rashidi, H. H., Pantanowitz, J., Hanna, M., Tafti, A. P., Sanghani,
P., Buchinsky, A., ... \& Pantanowitz, L. (2025). Introduction to
Artificial Intelligence (AI) and Machine Learning (ML) in Pathology \&
Medicine: Generative \& Non-Generative AI Basics. Modern Pathology,
100688.

177. Annamalai, N., Bervell, B., Mireku, D. O., \& Andoh, R. P. K.
(2025). Artificial intelligence in higher education: Modelling students'
motivation for continuous use of ChatGPT based on a modified
self-determination theory.~Computers and Education: Artificial
Intelligence,~8, 100346.

178. Rahimi, A. R., Sheyhkholeslami, M., \& Pour, A. M. (2025).
Uncovering personalized L2 motivation and self-regulation in
ChatGPT-assisted language learning: A hybrid PLS-SEM-ANN
approach.~Computers in Human Behavior Reports,~17, 100539.

179. Illangarathne, P., Jayasinghe, N., \& de Lima, A. D. (2024, May). A
Comprehensive Review of Transformer-Based Models: ChatGPT and Bard in
Focus. In 2024 7th International Conference on Artificial Intelligence
and Big Data (ICAIBD) (pp. 543-554). IEEE.

180. Qian, W., Arumugam, N., Shi, Q., Yang, J., \& Dong, M. Factors
Influencing Learning Adjustment of Vocational Undergraduates: An AI Chat
GPT Perspective using SmartPLS Model.

181. Makrygiannakis, M. A., Giannakopoulos, K., \& Kaklamanos, E. G.
(2024). Evidence-based potential of generative artificial intelligence
large language models in orthodontics: a comparative study of ChatGPT,
Google Bard, and Microsoft Bing. European Journal of Orthodontics,
cjae017.

\end{document}